\documentclass[graybox]{svmult}

\usepackage{mathptmx}       
\usepackage{helvet}         
\usepackage{courier}        
\usepackage{type1cm}        

\usepackage{amssymb}
\usepackage{makeidx}         
\usepackage{graphicx}        
\usepackage{multicol}        
\usepackage[bottom]{footmisc}
\makeindex             

\newcommand{\htwo}{\mathrm{H}_2}
\newcommand{\cc}{\mathrm{cm}^{-3}}
\newcommand{\kelvin}{\mathrm{K}}
\newcommand{\kb}{k_{\mathrm{B}}}
\newcommand{\zsun}{Z_{\odot}}
\newcommand{\msun}{M_{\odot}}
\newcommand{\racc}{r_{\mathrm{acc}}}
\newcommand{\au}{\mathrm{AU}}
\newcommand{\msunperyr}{M_{\odot}\,\mathrm{yr}^{-1}}

\begin{document}

\title*{Star Formation for Predictive Primordial Galaxy Formation}
\author{Milo\v s Milosavljevi\'c \and Chalence Safranek-Shrader}
\institute{Milo\v s Milosavljevi\'c \at University of Texas, Austin, TX, 78712, USA, \email{milos@astro.as.utexas.edu} \and
Chalence Safranek-Shrader \at University of California, Santa Cruz, CA 95064, USA, \email{ctss@ucsc.edu} }

\maketitle

\abstract*{The elegance of inflationary cosmology and cosmological perturbation theory ends with the formation of the first stars and galaxies, the initial sources of light that launched the phenomenologically rich process of cosmic reionization.  
Here 
we review the current understanding of early star formation, emphasizing 
unsolved problems and technical challenges. We begin with the first 
generation of stars to form after the Big Bang and trace how they 
 influenced subsequent star formation. The onset of 
chemical enrichment coincided with a sharp increase in the overall physical complexity 
of star forming systems. {\it Ab-initio} computational treatments are just now entering the 
domain of the predictive and are establishing contact with local observations of the relics of this ancient epoch. }

\abstract{
The elegance of inflationary cosmology and cosmological perturbation theory ends with the formation of the first stars and galaxies, the initial sources of light that launched the phenomenologically rich process of cosmic reionization.  
Here 
we review the current understanding of early star formation, emphasizing 
unsolved problems and technical challenges. We begin with the first 
generation of stars to form after the Big Bang and trace how they 
 influenced subsequent star formation. The onset of 
chemical enrichment coincided with a sharp increase in the overall physical complexity 
of star forming systems. {\it Ab-initio} computational treatments are just now entering the 
domain of the predictive and are establishing contact with local observations of the relics of this ancient epoch. }

\section{Introduction}
\label{sec:1}

Numerical modeling of early star formation is propelled by the optimism that simulations will facilitate a predictive theory of the young Universe, the theory that will be required to interpret observations of the epoch of reionization.  The motivation for the development of ever more accurate numerical simulations ranges from the fundamental to the phenomenological: 

(1) {\it Ab initio} simulations of metal-free star formation have successfully synthesized protostars from initial conditions realizing inflationary random fluctuations \cite{Yoshida06,Yoshida08,Greif12}.\footnote{In the loosely defined sense employed here, an `{\it ab initio}' model is one that approximates an astrophysical system with discretized equations of reactive, gravitating radiation-hydrodynamics, with initial conditions defined by the standard cosmological model, and relegating stellar evolution and nucleosynthesis to a subgrid model.}  This has brought into focus the prospect of extending such simulations to the earliest star clusters, galaxies, radiation backgrounds, and the intergalactic medium (IGM).  {\it Ab initio} treatments of star formation in the Galactic molecular clouds have begun to converge in the numerically synthesized stellar initial mass function (IMF), at least at low and intermediate stellar masses \cite{Bate12,Krumholz12,Bate14}.  A supporting development is recent progress in multidimensional direct simulation of explosive pair instability and core collapse supernovae \cite{Janka12,Chatzopoulos13,Chen14a,Chen14b,Couch14,Wongwathanarat14}.  These long-awaited breakthroughs anticipate a genuinely predictive theory of stellar nucleosynthesis, an essential piece in the puzzle of primordial star formation.  There is also the hope that a convergence of numerical investigations that bridge the gap between interstellar and stellar interior scales will reveal the true pathways to massive black holes that are realized in nature \cite{Volonteri10}.  In the case of massive black hole formation, the {\it ab initio} numerical approach is essential because the very first event horizons, which must have appeared concurrently with the first stars and galaxies, are unlikely to have produced directly-detectable radiative signatures.\footnote{The formation of a black hole may be detected as a gamma-ray burst (GRB), but in spite of decades of close scrutiny, we still lack a unique unifying model for the GRB phenomenon.}  Accelerating these efforts are new algorithms and software libraries, such as the moving mesh hydrodynamics \cite{Springel10,Duffell11,Yalinewich14} and higher order radiative transfer methods \cite{Davis12,Skinner13,Klassen14,Tsang15}.  There is also  continued progress on subgrid scale modeling required to attain ``large eddy" consistency of the simulations (i.e., the subgrid prescriptions that are required to correctly capture the unresolved scale dynamics so as to make the the model dynamics independent of numerical resolution) \cite{Schmidt14}.

(2) Schemes are maturing for extracting cosmological correlations from radiative signals from the epoch of reionization: the 21 cm neutral hydrogen hyperfine line \cite{Fan06,Furlanetto06,Morales10,Pritchard11} and the secondary anisotropy in the cosmic microwave background (CMB) due to the Sunyaev-Zel'dovich effect \cite{McQuinn05,Mesinger12,Park13}, and more speculatively, a cosmic infrared background anisotropy from reionization-era unresolved point sources \cite{Kashlinsky15,Helgason15}.  Since the radiation from stars and accreting stellar remnants modulates the measurable fluctuating backgrounds, the transfer functions mapping the statistics of inflationary perturbations onto spatial correlations in fluctuating backgrounds encode the multiscale astrophysics of star formation and stellar evolution. This is especially true during the early stages of reionization when the `astrophysical bias' (especially of the non-linear kind) is inevitably the strongest.  For example, the production and escape of Lyman continuum (LyC) and Lyman-$\alpha$ (Ly$\alpha$) radiation from ancient galaxies is, via the impact of photoionization and supernovae on the protogalactic medium, sensitive to the precise rhythm and spatial incidence of star formation.  The anticipation that star formation in the earliest galaxies will eventually become numerically tractable drives the confidence that mapping of the early reionization will unlock access to a wide range of inflationary wavelengths, including those that would have been washed out in lower-redshift correlations.  The secondary non-gaussianity imprinted during reionization, which reflects the physics of star and galaxy formation, is degenerate with the much sought-after non-gaussianity of inflationary origin \cite{Lidz13}.

(3) The inventory of reionization-epoch galaxy candidates detected via the Lyman break and other related techniques is growing rapidly.  Observations with the James Webb Space Telescope (JWST \cite{Gardner06}) and ground-based thirty meter class telescopes, possibly enhanced with gravitational lensing \cite{Maizy10,Zackrisson12,Richard14,Zackrisson14}, will enable a giant leap in the direct detection and spectroscopic confirmation of ancient galaxies, with detection limits reaching galactic stellar masses as low as $\sim10^5\,M_\odot$ \cite{Pawlik11,Pawlik13}.  The galaxy luminosity functions, if naively extrapolated at the faint end and combined with the CMB polarization measurement of the electron scattering optical depth to recombination \cite{Planck15}, indirectly indicate a significant contribution of hitherto undetected, low-mass galaxies to cosmic reionization \cite{Robertson10}. A faint-end turnover in the luminosity function would reflect a rapidly declining efficiency of galaxy formation; such a turnover is currently undetected.  Attempts are underway to precede the next-generation observations with theoretical predictions of the radiative signatures of the high-redshift sources.  The numerical techniques typically involve hybrids of {\it ab initio} treatments of gravitational and gas dynamics on spatial scales $\gg 1\,\mathrm{pc}$ and synthetic prescriptions for star formation and its effects on smaller scales \cite{Wise12,Agertz13,Pawlik13,Jeon14}.  High-resolution simulations combined with radiative transfer carried out in post-processing are essential because the equivalent widths of the Ly$\alpha$ line and other nebular lines are functions of the multiscale configuration of the star-forming complexes \cite{Dijkstra06,Hansen06,Verhamme06,Zheng10,Smith14}. Radiation pressure may even play a dynamical role in some cases, mandating a fully-coupled radiation-hydrodynamical treatment \cite{Milosavljevic09,Wise12b}.

(4) The nucleosynthetic products of early star formation, once expelled from star formation sites, polluted the IGM and the circumgalactic medium (CGM).  Among the absorption line systems in isolated, low-mass galaxies and the unvirialized cosmic web, there could exist those polluted by only the earliest stellar generations \cite{Becker12}, or even those containing completely pristine, unpolluted gas \cite{Fumagalli11}.  Transmission spectroscopy of such systems can provide direct access to primordial nucleosynthesis.  With this in mind, damped Ly$\alpha$ absorbers and Lyman limit systems are being searched for very iron-poor instances \cite{Cooke11}.  Here, again, it is the astrophysics---that of supernovae and their gaseous remnants, superbubbles, and fountains---that is the principal poorly determined theoretical element \cite{Salvadori12,Kulkarni13,Maio13,Kulkarni14,Webster15}.  The challenge lies in the multiscale statistical character of the observables: it will not suffice to qualitatively explain how metal-poor absorbers formed around localized nucleosynthetic sites, but one must also model their cosmic statistics.  The detailed mechanics of enrichment also critically determines under what conditions, and with what properties, do metal-free stellar systems continue to form in pristine pockets as metal-enriched star formation begins to dominate. The distinct stellar IMF and high photosopheric temperatures ($\sim10^5\,\mathrm{K}$) of metal-free stars make such systems particularly appealing epoch-of-reionization spectroscopic targets \cite{Sobral15}.

(5) Credible relics of primordial galaxies have been found in the Milky Way's dwarf satellite galaxy population.  The ultra-faint dwarf (UFD) satellite galaxies have stellar masses as low as $\sim 10^3\,M_\odot$ and dynamical mass-to-light ratios (within the stellar half-light radii) as high as $\sim 10^3\,M_\odot/L_\odot$ \cite{Simon07,Simon11,Willman11,Koposov15}. They also have stellar ages indistinguishable, within present uncertainties, from the age of the Universe \cite{Brown13,Brown14}, as well as metallicities lower than measured in any other stellar systems in the nearby universe \cite{Kirby11}.  The information contained in the UFD population is relatively rich. They manifest dark matter clustering on comoving wavelengths $\lesssim 1-10\,\textrm{kpc}$. Their orbits constrain the cosmic collapse redshift and the initial distance to the most massive progenitor of the Milky Way.  Disrupted UFD-type objects could have contributed to the assembly of the metal-poor Galactic stellar halo \cite{Kirby08,Norris10,Lai11}. Most interestingly, their stellar chemical abundances exhibit a huge variation not only between galaxies, but also within individual galaxies \cite{Vargas13}.    The aspiration that the diverse nucleosynthetic inventory returned to the universe through stellar mass loss and supernovae will be reverse-engineered from stellar abundance patterns, e.g., with principal component analysis \cite{Ting12}, comes the closest to feasibility in UFDs and the Galactic halo.  To ensure that such analyses deliver valid conclusions, hydrodynamic effects that skew the abundances in star-forming clouds relative to those of the contributing sources must be characterized with {\it ab initio}-type simulations \cite{Ritter14,Sluder15}.

The perspective of this review differs in its emphasis on the theoretical and computational bottlenecks from a majority of other reviews \cite{Bromm11,Bromm13,Greif14b} in the field of computational reionization-era star and galaxy formation.  We survey the elements of the theory with an emphasis on unsolved problems and numerical challenges.  
We admit a decidedly fine-grained, ground-up perspective in the narrative that follows.\footnote{In the interest of brevity, only an arbitrary selection of representative, mostly recent literature is cited.  No attempt is made to provide a historical pespective and attribute credit to the originators of the ideas and discoveries covered in the review.  References to the milestones in the development of the theory of early galaxy formation can be found in, e.g., Greif \cite{Greif14b}.} 

\section{Metal-free Systems}
\label{sec:2}

The environmental influence, or \emph{feedback}, from first-generation, metal-free stars (commonly referred to as Population III) drove the transformation of the young Universe.  Quantitative features of metal-free star formation of particular consequence include star formation efficiencies (SFEs), stellar mass statistics (the IMF and its potential variation with local conditions), dynamical evolution of star clusters (e.g., the velocities with which stars are ejected from nascent clusters), stellar initial rotation rates and rotational evolution, stellar multiplicity (the statistics of stellar and compact remnant binaries), the geometry and hydrodynamics of H\,II regions, the character and nucleosynthetic yields of the supernovae, and the nature of accretion onto the compact remnants left behind. 

Each of these features could, in principle, have an order-of-unity or larger effect on early reionization and our ability to investigate it with high-redshift observations. For example, metal-free stars produce an order-of-magnitude more ionizing photons per baryon than metal-enriched stars.  The escape of UV and X-ray photons from the lower mass halos (typically ``minihalos''\footnote{``Mini-halos" are those in which the compression by dark matter gravity alone does not heat gas to temperatures $\sim 10^4\,\mathrm{K}$ at which radiative energy loss to collisional excitation of the Ly$\alpha$ transition dominates thermodynamic evolution. The halos that do reach these critical temperatures are ``atomically-cooling halos".}) that host metal-free stars may be easier.  Simultaneously, the feedback from star formation on the gas content in lower-mass halos could be more deleterious. The physics and observational signatures of early reionization can also be especially sensitive to the details of metal-free star formation if metal-free stars are associated with unusual stellar evolutionary outcomes: extreme supernovae, gamma-ray bursts, and massive black holes.  

Ideally, the effects of metal free stars would be calibrated through direct observation.  However, just as metal-free stars may be difficult to observe directly as they form in the young Universe, their potential low-mass, long-lived relics are expected to be rare; their candidates have been hard to find in the Galactic spectroscopic surveys. Indeed, the signature of metal-free stars has not yet been unambiguously detected, even indirectly, in stellar chemical abundance patterns or the extragalactic background light.  But other promising indirect avenues remain and are the current focus of attention.  The first cosmic objects' unknown properties are encoded in the reionization epoch's large scale signatures which we hope to reverse-engineer from upcoming IGM observations. Anticipating observational probes, recent progress has been confined to the theoretical plane, and so in this section, we review elements of the theory of star formation and evolution preceding chemical enrichment.

\subsection{The Initial Gravitational Collapse}

Metal-free protogalactic gas clouds with temperatures $<10^4\,\mathrm{K}$ cool by ro-vibrational line emission from molecular hydrogen (H$_2$) \cite{Bromm04} and, under certain special circumstances, from hydrogen deuteride (HD) \cite{Johnson06,McGreer08}.  At temperatures $\sim 10^4\,\mathrm{K}$, radiative cooling is dominated by Ly$\alpha$ line emission (at densities below $10^6\,\mathrm{cm}^{-3}$) and free-bound emission from H$^-$ formation (at higher densities \cite{Omukai01}). Local and global ultraviolet (UV) and X-ray radiation may be capable of altering the thermodynamic trajectory of gravitationally collapsing clouds \cite{Omukai08,OShea08}. Sufficiently strong UV radiation in the narrow Lyman-Werner spectral range ($11.2-13.6\,\mathrm{eV}$) dissociates molecular hydrogen in metal-free gas.

Before delving into the sensitivity of the initial gravitational collapse to radiation backgrounds, it is worth noting that they can endow the dynamics of reionization fronts with a long-range, nonlinear coupling.  Sufficiently early on, metal-free stars and their metal-enriched descendants were the dominant sources of UV, and perhaps even X-ray, backgrounds (though the escape of UV backgrounds from metal-free star forming systems could be less than efficient \cite{Schauer15}).  The backgrounds in turn inhibited concurrent metal-free star formation at locations removed from the sources by modifying the conditions for star formation in metal free clouds. Radiation emitted from rare, high density cosmic peaks first suppressed metal-free star formation in moderate (and low) density environments, and then reionized these environments in their pristine, starless state.  Estimates of the star formation rate (SFR) and UV background intensity at redshifts $z\sim10-20$ diverge widely in their predictions \cite{Ahn09,Holzbauer12,Fialkov13,Dijkstra14}, unsurprising given that {\it ab initio} estimates of the SFEs and IMFs, which figure as key model parameters, have not yet converged.

Critical for the formation of metal-free stars is how thermodynamic evolution in the gravitational potential of a dark matter halo escorts the gas to the Bonnor-Ebert threshold whereupon it can begin runaway gravitational collapse.  Numerical results suggest that the mass infall rate at the beginning of runaway collapse has a strong effect on the fragmentation character and the end point of the collapse \cite{Kratter11,Vorobyov13}. The external radiation field intensity influences the mass infall rate through its effect on the mass and temperature of the initial self-gravitating gas cloud.  The mass infall rate from an initially marginally Jeans-unstable cloud with velocity dispersion $\sigma$ (representing a combination in quadrature of the thermal and turbulent dispersion) is $\sim \sigma^3/G$.  Thus we expect the infall rate to increase as we consider, in a sequence of increasing $\sigma$, an initially cold neutral (molecule-cooled), warm neutral, ionized, and supersonically turbulent warm cloud (for illustration, e.g., $\sigma\sim (1,\,5,\,10,\,20)\,\mathrm{km}\,\mathrm{s}^{-1}$, respectively).  If $\sigma$ remains approximately constant as the density increases, the collapse proceeds in a self-similar fashion, with the density profile rising toward the center as $\rho(r) \propto\sigma^2/(Gr^2)$ up to a maximum density. In the following we summarize the three scenarios for star formation in metal-free gas.

\begin{itemize}
\item If the intergalactic molecule-photodissociating background does not sufficiently diminish the H$_2$ abundance in a bound cosmological object, molecular cooling reverses the gas temperature rise induced by compression when temperatures and densities respectively reach $\sim 10^3\,\mathrm{K}$ and $\sim 1\,\mathrm{cm}^{-3}$ \cite{Abel00,Bromm01}. This typically occurs in a dark matter halos with virial mass $\sim 10^5-10^6\,M_\odot$ (the critical halo mass increases with decreasing redshift \cite{OShea07}).  The temperature decreases by a factor of a few by the point the density reaches $\sim 10^3\,\mathrm{cm}^{-3}$. Then the temperature rises again as the H$_2$ cooling transitions thermalize at $\sim 10^4\,\mathrm{cm}^{-3}$ and eventually become optically thick at $\sim10^9\,\mathrm{cm}^{-3}$.   Coincidentally, at approximately the same densities as at the onset of optical thickness, gravitational collapse encounters the centrifugal barrier \cite{Stacy10}, forcing the gas to circularize around a small collapsed core.  The orbiting gas develops non-axisymmetric perturbations and gravitational torques acting on these disturbances transport angular momentum outward.  Meanwhile, the central core collapses unimpeded to progressively higher densities.  Eventually, the density becomes so high, typically $\sim 10^{16}\,\textrm{cm}^{-3}$, that the gas begins trapping its own cooling radiation.  The radiation-trapping entity can be identified with a hydrostatically-supported protostellar core \cite{Yoshida06,Greif14,Hartwig14}.  Mass infall rates at sufficiently large radii, where the gas flow is unaffected by the centrifugal barrier, peak around $\sim 0.01\,M_\odot\,\mathrm{yr}^{-1}$, and subsequently decline \cite{OShea07,Stacy14}.

\item A strong non-ionizing UV background can suppress molecule formation and destroy preexisting molecules (above $11.2\,\mathrm{eV}$, by directly dissociating H$_2$ via the Solomon process \cite{Stecher67} and below that energy, by dissociating the H$_2$ formation catalyst H$^-$).  In such a background, runaway gravitational collapse is delayed until the halo mass grows to $\gtrsim 10^7\,M_\odot$ and compression by dark matter gravity heats the gas to $\sim 10^4\,\mathrm{K}$ \cite{OShea08,Omukai08}. The temperature plateaus there because the rate of energy loss to collisional Ly$\alpha$ line excitation increases steeply with temperature.  If the background is not too intense, the collapsing cloud starts self-shielding from the dissociating background at densities $\sim10-10^4\,\mathrm{cm}^{-3}$ (the critical density increases with increasing background intensity \cite{WolcottGreen11}) and cools approximately isobaricially to $\sim 400\,\mathrm{K}$ \cite{SafranekShrader12,Inayoshi14,Fernandez14,Latif14,Regan14}.  The subsequent thermodynamic evolution is similar to the background-free case, except that the mass infall rate is now an order-of-magnitude higher, peaking at $\sim 0.1\,M_\odot\,\mathrm{yr}^{-1}$.  However if a proximate, vigorously star forming, and possibly more massive halo is providing an intense molecule-dissociating flux  \cite{Bromm03b,Shang10,Agarwal14,Dijkstra14,Visbal14}, e.g., with an intensity $J\gtrsim 10^{-18}\,\mathrm{erg}\,\mathrm{s}^{-1}\,\mathrm{cm}^{-2}\,\mathrm{sr}^{-1}\,\mathrm{Hz}^{-1}$ \cite{Latif14,Regan14}, or perhaps an order of magnitude higher \cite{Latif15}, the collapse avoids the molecular-cooling-dominated evolutionary phase altogether.  If this happens in a relatively massive halo, the mass infall rate may be sufficiently high to create a supermassive protostar. Such an object could collapse directly into a massive black hole (more about this in \S\ref{sec:protostellar_evolution} and \S\ref{sec:black_holes} below).  

\item In externally ionized halos with virial masses $\gtrsim 10^8\,M_\odot$, gas simultaneously commences quasi-isothermal gravitational collapse and starts shielding itself from ionizing radiation at densities $\sim 0.1-1\,\mathrm{cm}^{-3}$. Rapid inside-out recombination, molecule formation, and perhaps even thermal instability ensue \cite{Johnson14}.  To date, the subsequent hydrodynamic evolution to protostellar densities has not been simulated in three dimensions. It could be similar to the H$_2$-suppressed case, potentially with somewhat higher turbulent mach numbers and mass infall rates, and could represent the pathway to the most massive star clusters that can form from metal-free gas. The requirements for simulating this process include photon-conserving, multi-source (or diffuse) ionizing radiation transfer and a direct integration of a chemical reaction network tracking the formation of H$_2$ out of chemical equilibrium.  This scenario for metal-free star formation could be realized even in a completely reionized Universe, perhaps even at low redshifts in dwarf-galaxy-type halos with masses $\sim10^9\,M_\odot$.
\end{itemize}

Common among the three versions of metal-free runaway gravitational collapse identified above are the following: an initial uninterrupted, quasi-isothermal collapse of gas, with a small gas mass, $\sim 0.1\,M_\odot$, first reaching protostellar densities, a relatively immediate activation of the centrifugal barrier after which the central protostar accretes through a protostellar disk with an initial radius $\sim 10\,\mathrm{AU}$, an increase of the radial extent of the disk as the protostar grows, and an intermittent gravitational instability in the disk that produces additional accreting protostars.  

\subsection{The Protostellar Disk}

After only a few years from the formation of the first protostar, the metal-free star-forming system can be formally described as a supersonic compressible flow coupled to $N$ gravitating, accreting,  radiating, and potentially even mass-losing bodies.  If radiation from the protostars does not entirely reverse the accretion, the protostars can continue to grow for $\sim \mathcal{O}(1\,\mathrm{Myr})$, until a central crisis---catastrophic central gravitational collapse or thermonuclear ignition---destroys the star.  With orbital time scales as short as $\sim \mathcal{O}(1\,\mathrm{yr})$, the system could evolve through $\sim\mathcal{O}(10^6)$ revolutions before the protostars stop growing. With to-date numerically-convergent simulations of protostellar disks extending for only $\sim\mathcal{O}(10\,\mathrm{yr})$ \cite{Greif12,Becerra15}, integration over a physically realistic number of revolutions is well beyond the current numerical capabilities.  

The radius of the fragmenting disk increases as the primary protostar grows massive, e.g., to $\sim10^4\,\mathrm{AU}$ in $\sim 0.3\,\mathrm{Myr}$ \cite{SafranekShrader12}.  This is because the disk extends only to radii at which angular momentum transport by gravitational torques is not so rapid as to force the matter to move inward on a few orbital times.  Angular-momentum-transporting disturbances such as spiral arms are suppressed (stabilized) by the tidal field of the central point mass, here a protostar or a compact group of protostars.  

How satellite bodies embedded in a self-gravitating gaseous disk grow remains an open theoretical problem even in metal-enriched star and planet formation.  Early results suggest knife-edge parameter sensitivity and the potential for chaotic dynamical interaction among the satellite bodies \cite{Kratter10,Vorobyov10,Vorobyov13}.  The mass growth of individual protostars can be stunted by many-body ejection from the natal gas cloud \cite{Greif11,Greif12}, by photoionization of the gas if any of the massive protostars becomes a strong source of ionizing radiation (which normally requires contraction onto the main sequence) \cite{Hosokawa11,Stacy12,Hirano14,Susa14} or by protostellar mass loss.  For none of these processes is the existing theory predictive.  Why?

When we attempt to simulate such a system, we are hindered by the susceptibility of self-gravitating gas dynamics to develop subtle numerical artifacts if the Jeans length $\lambda_{\rm J}$ is not resolved by a sufficient number of mesh elements of size $\Delta x$ \cite{Truelove97}.  One typically requires that $N_{\rm J} =\lambda_{\rm J}/\Delta x$ be greater than a minimum such ratio, usually estimated at $N_{\rm J,min}\gtrsim  32$.\footnote{This empirical resolution requirement expressed in terms of $N_{\rm J,min}$ seems to arise from a combination of a genuine requirement to capture gravitational instability \cite{Truelove97}, and a wish to resolve cascading turbulent fluctuations on scales of the gravitationally unstable wavelengths \cite{Federrath11,Turk12}. With the latter requirement in mind, it is not clear that the customary mesh refinement criterion solely based on the local Jeans length is entirely adequate.}  In a gas disk on the verge of local gravitational instability with a stability parameter $Q\sim 1$, the Jeans length is of the order of twice the disk scale height $\lambda_{\rm J}\sim 2h$, and the scale height is related to the orbital Mach number via $h\sim R/\mathcal{M}$.  The total number of mesh elements required to simulate this disk in three dimensions is $\sim 2\pi h R^2 /\Delta x^3$. If the orbital Mach number  is moderately large, $\mathcal{M}\gtrsim \mathcal{O}(10)$, the number of mesh elements required is $\sim N_{\rm J,min}^3 \mathcal{M}^2\gtrsim \mathcal{O}(10^7)$. Additional resolution elements must be introduced around any secondary protostars in the disk that will form their own smaller protostellar disks feeding from their corotation annuli in the primary's disk.  If this system is to be integrated over $\sim \mathcal{O}(10^6)$ revolutions each requiring $\sim100$ time steps per revolution, the requirements are well outside of what is presently feasible.  Such ambitious simulations would have to conserve angular momentum over the entire evolution if they are to attain convergence in the physical observables. With most computational hydrodynamic schemes, this would present a challenge in itself.

In complexly organized, supersonic, self-gravitating, and differentially rotating flows, the moving mesh methods mentioned in \S~\ref{sec:1} are already more efficient and accurate than Eulerian methods with adaptive mesh refinement \cite{Duffell12}.  We need a deeper mathematical understanding of the origin of finite resolution effects in self-gravitating simulations, aiming, for example, to improve the discretization of the gravitational source term so that $N_{\rm J,min}$ can be reduced.  The recent introduction of a Discontinuous Galerkin scheme for astrophysical hydrodynamics \cite{Schaal15} should pave the way to treating the hyperbolic hydrodynamic subsystem and the elliptic gravitational Poisson subsystem on equal footing. We also need further testing and fine-tuning of the subgrid prescriptions (`sink particles') that are used to coarse-grain unresolved sites of gravitationally-collapse.  The flows feeding real protostars and their disks are supersonic and that renders the physical accretion rate independent of the properties (e.g., size) of the accreting object, as in the idealized Bondi-Hoyle-Lyttleton problem.  Sink particles, however, typically coarse-grain the flow on length scales on which accretion is subsonic; then, the artifacts of coarse graining can propagate outward and compromise the physical self-consistency of the flow.  With such potentially serious systematic effects in mind, numerical convergence must be sought systematically, by judiciously scaling physical and numerical parameters (accretion rates, cooling functions, opacities, subgrid prescriptions, etc.) into ranges that render the evolution of the protostellar disk computable over its entire lifetime, and extrapolating when necessary toward the astrophysical limit.  
 
When a massive protostar contracts onto the main sequence it begins to emit ionizing radiation.  This radiation gradually evaporates the mass inflow and eventually terminates further accretion onto the protostar.  Before the object contracts onto the main sequence, some mechanical energy dissipated by the accreting matter and emitted as ``accretion luminosity" can be re-absorbed in the accretion flow and that can suppress or delay fragmentation \cite{Smith11,Stacy12}.  These processes have so far been simulated only in idealized, two-dimensional geometries \cite{Hosokawa11}, or at relatively low resolution \cite{Stacy12}.   Photon-conserving ionizing radiative transport schemes in which every ionizing photon is either explicitly absorbed by the gas, or leaves the computational domain, may still be susceptible to artifacts created by not resolving the thin photo-evaporation flows that are associated with D-type ionization fronts \cite{Bertoldi90}.

\subsection{Protostellar Evolution at Low and High Accretion Rates}
\label{sec:protostellar_evolution}

The internal evolution of growing protostars can now be readily computed with stellar evolutionary codes such as \textit{Modules for Experiments in Stellar Astrophysics} (MESA) \cite{Paxton11,Paxton13}. One can then explore the sensitivity of the protostellar evolution to the character of mass accretion, specifically whether material arrives at the surface through radially directed supersonic accretion, through an accretion disk, or through smaller, satellite protostars that migrate in the disk and merge with the primary.  Stellar merging can be modeled with stellar evolutionary codes provided one first quantifies, ideally with {\it ab initio} simulations, the degree to which merging stars get mixed in the merger. Compared to full mixing, the opposite extreme is ``entropy sorting" that preserves the stratification of the progenitors. Entropy sorting constructs a merged stellar model by combining (locally mixing) progenitor shells with equal specific entropies, such that new entropy is not generated in the merger. When lower-mass protostars merge with a more massive protostar, they deliver mass to the latter at a lower entropy than if the accretion were via vertical or disk accretion.  This can influence protostellar evolution, and can perhaps even make it easier for the protostar to become an ionizing radiation source.  Protostellar merging in the presence of continued accretion has not been examined in the metal-free regime.  

Stellar evolutionary calculations show that at relatively low protostellar accretion rates $\lesssim 0.01\,M_\odot\,\mathrm{yr}^{-1}$, metal-free protostars contract onto the main sequence upon reaching a mass of roughly $M_{\rm MS} \approx 30\,M_\odot (\dot{M}/10^{-3}\,M_\odot\,\mathrm{yr}^{-1})^{1/3}$ \cite{Hosokawa09,Hosokawa10}. The Kelvin-Helmholtz contraction is preceded with a swelling of the radius to $10-1000\,R_\odot$ (increasing with increasing accretion rate) as the core opacity drops and an entropy peak travels toward increasing mass coordinates.  At higher accretion rates $\gg 0.01\,M_\odot\,\mathrm{yr}^{-1}$, the evolution changes qualitatively, with material settling onto the surface always ahead of the entropy peak.  The star eludes contraction, with the radius remaining swollen at $R\sim 1000\,R_\odot$ as the stellar mass surpasses the maximum of $\sim200\,M_\odot$ seen in the nearby Universe (at nonzero metallicities) \cite{Hosokawa12a,Hosokawa13}.  Such massive, rapidly accreting protostars are luminous; they radiate near the Eddington limit, but are too cool at the surface ($T_{\rm eff}\lesssim 10^4\,\mathrm{K}$) to be appreciable ionizing sources.  Hosokawa et al.\ have noted that by not photoionizing its surroundings, the protostar does not subvert its own mass supply and might continue growing essentially arbitrarily large so that it could eventually collapse into a massive black hole.  For this scenario to be realized, the secondary satellite protostars that may be forming in the primary's protostellar disk must also not become ionizing sources with their own H\,II regions.  Also, the accretion must remain sufficiently steady on the protostar's thermal relaxation time scale; otherwise, contraction onto the main sequence and deleterious photoionization of the circumstellar environment can take place in between accretion rate bursts \cite{Sakurai15}.

Does the conjecture that supermassive stars form in objects with suppressed molecular cooling withstand closer theoretical scrutiny?  The rapidly accreting massive protostar is centrally supported by radiation pressure.  Hydrostatic stellar evolutionary models like those of Hosokawa et al.\ are extremely superadiabatic near the surface (specific entropy decreases precipitously with increasing radius).  The velocity of convective overturn, at least as prescribed by the mixing-length theory, is transsonic or even supersonic.  This is similar to asymptotic giant branch (AGB) stars and begs the question of stability and mass loss \cite{Inayoshi13}.  An interesting feature is the surface density inversion, which is ubiquitous in hydrostatic models of AGB stars and of rapidly accreting massive protostars, supermassive stars, and quasi-stars (rotating supermassive protostars in which the center has collapsed into a black hole but the envelope is still in place) \cite{Begelman08,Yungelson08,Ball11}.  Direct numerical simulation of near-Eddington, convecting envelopes has so far only been attempted in AGB stars and curiously so, ignoring the radiation pressure force \cite{Freytag08,Chiavassa14}.  The upper envelope is intensely turbulent and exhibits dynamical radial oscillations.  Density inversions appear, but then the cold, surface layer subducts, perhaps on a free-fall time, deep into the underlying higher-entropy layer, a behavior not captured in hydrostatic models but recently brought into focus by Brandenburg \cite{Brandenburg15}.  All this is important because the computed structure and evolution of rapidly accreting massive protostars may turn out to depend sensitively on the surface boundary condition.\footnote{The standard boundary condition employed in stellar-evolutionary calculations could be far from the true, time-averaged structure of the surface. For example, the envelope structure may depend on how extremely superadiabatic is the surface gradient allowed to be. }

Short of global multidimensional radiation-hydrodynamical simulations of inflated protostellar envelopes, the possibility of a yet unaccounted for instability that would disrupt the envelope cannot be excluded.  Turbulent momentum flux (i.e., turbulent pressure) associated with convection clearly contributes to the radial force balance near the surface but is typically not taken into account in quasi-hydrostatic structural models.  As the energy flux carried by convection increases, quasi-hydrostatic solutions may no longer exist and a supersonic outflow may develop \cite{Blandford04}.  At present, however, the formation of a supermassive, metal-free protostar remains a possibility; its descendant, as we discuss in \S~\ref{sec:black_holes}, could be a massive black hole.

\subsection{Stellar Evolution with Rotation}

Very massive metal-free stars ($\gtrsim 100\,M_\odot$) that have contracted onto the main sequence have convective interiors.\footnote{See recent reviews of the structure and evolution of rotating, massive, metal-free stars by Maeder \& Meynet \cite{Maeder12} and Hirschi \cite{Hirschi15}.}  The convective core contains a large fraction of the stellar mass and is chemically homogeneous.  If acceleration by radiation pressure on metal lines is the main stellar mass loss driving mechanism, the mass loss rate is a decreasing (e.g., an approximately inverse) function of the surface metallicity.  In the zero metallicity limit, line-driven mass loss becomes negligible.  However, proximity to the Eddington limit opens the possibility of wind driving by continuum radiation pressure, perhaps from only over a portion of the stellar surface (e.g., the poles).  Internal circulation and convection dredge up hydrostatically-synthesized nucleosynthetic products from the core to the surface and that can potentially increase the mass loss rate.  If the protostar is fed from an accretion disk, it should rotate rapidly.  At present it is uncertain if differential rotation, e.g., at the star-disk interface, can amplify the magnetic field enough to launch a magnetic-braking wind \cite{Tan04}.  

In a rapidly rotating star, meridional circulation, convection, and a variety of hypothesized magnetohydrodynamic instabilities transport chemical elements and angular momentum in the stellar interior.   Some of these processes are advective, transporting angular momentum either inward or outward; others are diffusive, acting to smooth out angular velocity gradients.    {\it Ab-initio} modeling of stellar interiors is challenging due to the vast disparity of time scales \cite{Glatzmaier13}: dynamical, shear, convective (Brunt-V\"ais\"al\"a), radiative (Kelvin-Helmholtz), circulation (Eddington-Sweet), etc.  Even in two dimensions, the modeling is challenging because the various transport processes are sensitive to tiny gradients that must be computed to high accuracy.  

Some aspects of stellar evolution are robust, e.g., the duration of the hydrogen-burning phase  ($\sim 2-3.5\,\mathrm{Myr}$) depends very weakly on stellar mass and rotation.  Other aspects of the physics of rotating stellar interiors remain unsettled. For example, how much is magnetic field amplified in radiative zones? What is the saturation level of the radial field component to which the angular-momentum-transporting Maxwell stress is proportional?  A pervasive amplification of the radial field would rotationally couple the surface to the core with consequences for how the core would eventually collapse, for the formation and character of the compact remnant, and for nucleosynthesis.  How significant is the post-main-sequence spin down?  It is standard at present to execute stellar evolutionary calculations with hypothetical, uncalibrated prescriptions for angular momentum transport coefficients, such as a prescription modeling the conjectured Tayler-Spruit dynamo.   Global magnetohydrodynamic simulations of differentially rotating stars seem necessary to test and lend credence to the prescriptions.

\subsection{Black Holes and Explosions}
\label{sec:black_holes}

The landscape of stellar evolutionary outcomes has been mapped out as a function of metallicity \cite{Heger03}, and at zero metallicity, as a function of the rotation rate \cite{Yoon12}.  The central assumption entering stellar evolutionary calculations from which outcomes are derived is that stars start their evolution on the zero-age main sequence and then evolve in isolation, not receiving further mass input through accretion or stellar merging (the validity of this assumption remains to be tested with {\it ab-initio} methods). Metal-free stars with masses $\lesssim 25\,M_\odot$ explode as core-collapse supernovae and leave behind neutron star remnants.  Above this critical mass but below $55-75\,M_\odot$ (the cutoff mass decreasing with increasing rotation rate \cite{Chatzopoulos12}),\footnote{In recognition of the theoretical uncertainties entering the computations cited here, we round off to the nearest multiple of $5\,M_\odot$} the stars collapse into black holes, either via fallback onto a proto-neutron star, or at higher masses, directly.  Stars more massive than this but with masses $\lesssim 85-110\,M_\odot$ produce pulsational pair instability core-collapse supernovae, leaving behind black hole remnants, and further with masses $\lesssim 190-240\,M_\odot$, explode as thermonuclear pair instability supernovae leaving no remnants (in both cases, again, the cutoff decreasing with increasing rotation rate \cite{Chatzopoulos12}).  The cores of still more massive stars collapse directly into black holes via a general-relativistic radial instability.  At very high stellar masses ($\sim 5\times10^4\,M_\odot$), exotic hyper-energetic thermonuclear explosions may be possible \cite{Chen14b}.

There are at least three aspects where the astrophysics of black hole formation in stellar core collapse is currently not predictive. Each, just like the protostellar disk evolution discussed in \S\ref{sec:protostellar_evolution}, is woefully beyond the range of applicability of low-dimensional models.  
\begin{itemize}
\item Core convection in the years and days preceding core collapse can be nonlinear, unstable, and so intense as to be dynamical, with the sonic, buoyancy, and nuclear time scales becoming comparable \cite{Arnett11,SmithN14,Couch15}.  Giant convective cells can develop and strongly break the spherical symmetry of the pre-supernova state of the star.  And all this precedes the post-collapse instabilities such as the standing accretion shock instability (SASI).  The asymmetries may affect the dynamics in the crucial phase in which neutrinos from the proto-neutron star deposit energy in the infalling stellar envelope to potentially reverse infall and drive an explosion. 
\item If the ``delayed-neutrino" or another mechanism does explode the envelope, the nature and final mass of the compact remnant depends on the magnitude of fallback, namely, how much mass either evades initial outward acceleration, or upon initial ejection turns around and falls back toward the remnant.  This process can be strongly asymmetric, e.g., in the presence of an initial asymmetry in the envelope (from pre-core-collapse evolution) or from aspherical instabilities in the explosion itself.  The innermost mass element of the ejected envelope may gravitationally pull the compact remnant in one direction, assisting fallback in that direction and imparting an impulse to the remnant \cite{Janka13}.  A birth kick of even $\sim 100\,\mathrm{km}\,\mathrm{s}^{-1}$, only one percent of the supernova ejecta velocity, could displace the remnant from the dense gas of the progenitor star's birth environment, thus interfering with its accretion from the interstellar medium (ISM).
\item At least some mass shells of the progenitor star may be rotating rapidly enough for the core collapse and its aftermath to be modified relative to the non-rotating case.  Differential rotation in the collapsing mass elements amplifies the stellar magnetic field.  The proto-neutron star or the black hole (whichever forms) will be endowed with a magnetosphere. Under certain conditions related to stellar stratification, rotation, and magnetization structure, a magnetohydrodynamic jet may develop.  Mechanical action of the jet transfers rotational energy from the central compact object to the envelope, potentially energizing the envelope enough to explode it.  This is the conjectured mechanism behind the super-energetic ($\sim10^{52}\,\mathrm{erg}$) ``hypernova" explosions.  High angular momentum mass elements may circularize in an orbit around the compact object, forming an accretion torus (or disk).  At very high accretion rates, the torus cools by neutrino emission and the compact object grows rapidly.  However at lower accretion rates, e.g., $\dot M\lesssim 0.1\,M_\odot\,\mathrm{s}^{-1}$, the neutrino-mediated energy loss is ineffective and the entropy generated by the dissipation of magnetohydrodynamic turbulence in the torus builds up enough to stop the infall, and perhaps even unbind the envelope \cite{Lindner12,Milosavljevic12}.  If a massive (or very massive) progenitor is rapidly rotating, it cannot be taken for granted that the entire envelope will accrete onto the black hole into which the core has collapsed. 
\end{itemize}

This last theoretical uncertainty is of particular concern for the quasi-star hypothesis for massive black hole formation, in which the core of a very massive, rotating, accreting protostar collapses into a black hole, but a much larger gas mass remains initially outside the black hole.  The gas is gravitationally bound and accretes on a time scale self-regulated to keep the outward energy flux below the Eddington limit at the surface of the cloud \cite{Begelman06,Begelman08}.  Configurations resembling quasi-stars may be generic outcomes of the scenario in which rapid central accretion in molecular-cooling-suppressed, metal-free halos builds a very massive protostar and ultimately a black hole.  {\it Ab-initio} simulations will tell if the black hole grows in a quasi-star-like fashion, or perhaps if a sudden energy transfer immediately after the black hole forms disrupts the envelope, stripping the black hole of its gas.

In black-hole-producing supernova explosions, the crippling theoretical uncertainty affecting predictions of the nucleosynthetic output relates to the mass cut dividing matter that is ejected from that which is consumed by the black hole. For example, all or none or any other fraction of, say, hydrostatically-synthesized oxygen, may be ejected, all depending on the fraction of the carbon/oxygen shell that has been consumed by the black hole.  The mass cut is not robustly predicted by spherical models and is usually left as a tunable free parameter \cite{Heger10}.  The mass cut approximation is itself of limited validity given the likely anisotropic character of the explosion with, e.g., nickel fingers penetrating the carbon/oxygen shell \cite{Wongwathanarat14}.

The anticipated high sensitivity of 21 cm observations of the epoch of reionization has compelled the consideration of even less theoretically determined end-of-life outcomes in metal free stars.  Their young supernova remnants are themselves sources of hard radiation backgrounds \cite{Johnson11}.  If metal free stars are in binaries that evolve into high-mass X-ray binaries---a highly uncertain proposition---they are another source of hard radiation that the IGM and affects both the sky-average and fluctuating 21 cm signal \cite{Mirabel11,McQuinn12,Mesinger13,Xu14,Ahn15,Watkinson15}.

\section{Metal-enriched Systems}
\label{sec:3}

The first elements heavier than lithium were synthesized in metal-free stars and were dispersed by their supernovae and non-explosive stellar mass loss.  Upon injection into the Universe, metallic atoms, molecules, and dust grains largely controlled the chemical and thermodynamic behavior of the enriched gas as it underwent gravitational collapse from intergalactic all the way to protostellar densities.  This introduced new dimensions of complexity not seen in metal-free star formation.  The increase in complexity is usually met with the idea that microscopic details of the star formation processes can be coarse-grained and replaced with macroscopic, effective models. The models of star formation used in the global modeling of reionization describe how the SFE, IMF, the initial mass function of star clusters, the frequency of X-ray binaries, etc., scale with parameters such as gas surface density, metallicity and dust content, pressure, turbulent Mach number, external radiation field, redshift, etc.  Theoretical predictions of signals from the epoch of reionization, however sophisticated in their treatment of cosmic fluctuations and cosmological radiative transfer, are thus invariably built on specific assumptions about the form of these effective scalings.  Empirical foothold on the scalings can in principle be gained from multiwavelength spectroscopic observations at high redshift, but typically, there are degeneracies, such as related to whether the hard radiation exciting the He\,$1640\,\mathrm{\AA}$ emission line is from hot stellar photospheres (indicating a metal-free population) or from accreting compact objects (see, e.g., Pallottini et al.~\cite{Pallottini15}).

Clues can also be sought in the local universe, both in the candidate fossils of the reionization epoch, and in what are considered to be nearby analogs of reionization-era galaxies.  Exploiting the information encoded in the fossils, stellar archaeological investigations attempt to relate chemical abundance patterns in the atmospheres of stars in the Milky Way's metal-poor stellar halo and dwarf satellite galaxies---potential relics from the early Universe---to the abundances expected from a diversity of contributing sources \cite{Frebel15}. This is challenging not only because the theoretical chemical yields are uncertain, but also because the metals' trajectory from injection until re-incorporation in stars is, as we shall see, exceptionally complex (see \S~\ref{sec:enrichment} below).  The local analogs of reionization-era galaxies (or, at least, local star forming complexes that are analogs of such complexes present during the reionization epoch) are identified through having young stellar populations and low or vanishing metallicities in H\,II regions \cite{Papaderos12,Annibali13}.\footnote{Whereas the local analogs can teach us about the star formation process under circumstances characteristic of the reionization epoch, it is not entirely clear if the analogy can be extended \cite{Borthakur14,Alexandroff15} to learning how star formation sources radiation backgrounds at high redshift.  The complication is that the circumgalactic medium of local galaxies that separates the star formation from the IGM may differ from its high-redshift counterpart in terms of its transmission properties.}  In these systems, the challenge lies in how to reconstruct the conditions that had precipitated what is selected on---what were the physical conditions that triggered the observed outburst of star formation?  More generally, what is the appropriate length scale on which the physical conditions should be coarse-grained (averaged) for the coarse-grained local environmental parameters to be the best predictors of star formation \cite{Kennicutt12}?  In view of this, we believe that a bottom-up, numerically-convergent understanding of metal-enriched star formation is mandatory to lend credence to programs predicting global signals from the epoch of reionization.  Like before, we limit the presentation to ideas derived from numerical work, but unlike for metal-free star formation, here we must reckon with a still small but growing body of observational constraints.

\subsection{Thermodynamic Impact of Metals}

Metals and dust grains provide a large number of efficient radiative cooling channels, much larger than accessible to the primordial pure hydrogen and helium gas. Secondary effects, such as dust grain catalysis of $\htwo$ formation and a sharp increase in opacity due to dust, are also significant in certain regimes. We begin with the interplay of chemistry, thermodynamics,  and self-gravity in a single isolated pre-stellar core undergoing gravitational collapse.

\begin{figure}[t]
\begin{center}
\includegraphics[width=\textwidth]{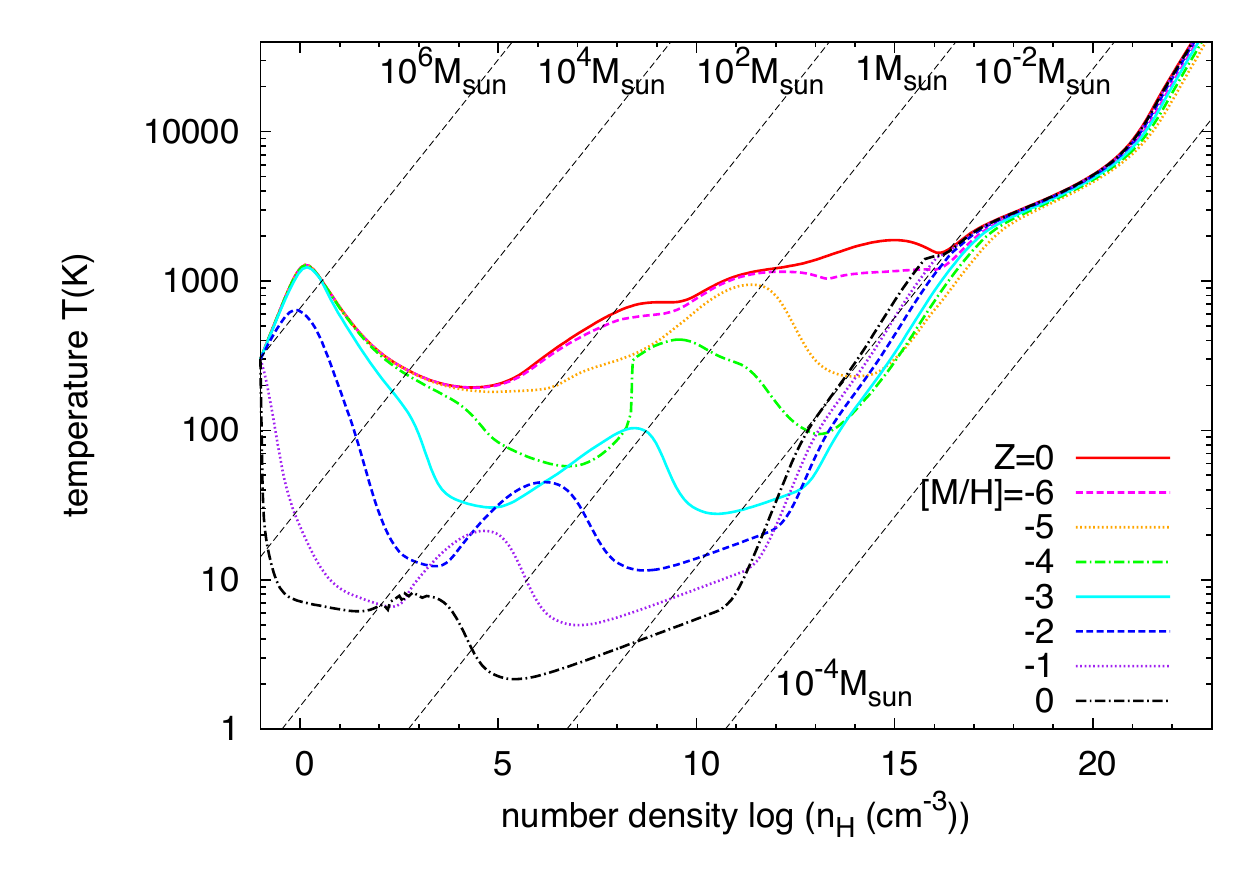}
\end{center}
\caption{
Density-temperature relationship for gravitationally collapsing dense cloud cores of varying metallicity, ranging from metal-free ($Z=0$; top, red solid line) to solar ($[{\rm M}/{\rm H}]=0$; bottom, black dash-dot line). Density is evolved in time according to $d\rho/dt = \rho / t_{\rm ff} \propto \rho^{3/2}$, a scaling approximating collapse in free fall.  At low densities the thermodynamic trajectories depend strongly on metallicity, such that the typical gas temperature decreases with increasing metallicity due to a combination of metal line emission and dust grain thermal emission. At high densities, $n\gtrsim10^{12}-10^{16}\,{\rm cm}^{-3}$, the trajectories converge when the continuum optical depth exceeds unity and a quasi-hydrostatic pre-stellar core has formed. Diagonal dashed lines are loci of constant Jeans mass as labeled in the plot (reproduced with permission from Omukai et al.\ \cite{Omukai10}).
}
\label{fig:Omukai}    
\end{figure}

Numerous studies have reported one-zone models tracking the coupled chemical and thermodynamical evolution of gas elements assumed to be located at the centers of gravitationally-collapsing clouds. While these simplified models exclude many effects influencing gas dynamics, such as magnetic fields, radiative feedback, and turbulence, they usefully highlight the thermodynamic processes that are relevant to the formation of prestellar cores, the evolutionary phase in which gas collapses and compresses from ISM ($n\sim0.1\,\cc$) to hydrostatic core ($n\sim10^{20}\,\cc$) densities. Figure \ref{fig:Omukai} shows the result of one such single-zone calculation by Omukai et al.\ \cite{Omukai10} that illustrates the evolution of central temperature as a function of central density for gravitationally collapsing pre-stellar cores of varying metallicity, ranging from zero to solar.  

As we have seen in Section \ref{sec:2}, the thermodynamic evolution of metal-free, pure H and He gas is regulated by the physics of the $\htwo$ molecule, the most effective coolant in neutral metal-free gas. Unless conditions are favorable for the formation of hydrogen-deuteride (HD), such as if there is an amplified free electron fraction \cite{Johnson06,McGreer08}, metal-free gas cannot cool below $T\sim100\,\kelvin$ owing to the relatively large energy difference, $E/\kb\approx500\,\kelvin$, between the lowest lying rotational transitions of $\htwo$. Second-order chemical and thermodynamic effects become important in certain phases of the collapse.  Among them, three-body reactions convert atomic gas to predominantly molecular gas while depositing latent heat from $\htwo$ formation in the gas. The $\htwo$ ro-vibrational lines become optically thick around $n\sim10^{10}\,\cc$. Collisionally induced emission (CIE) from $\htwo$ briefly becomes significant at $n\sim10^{15}\,\cc$ before this cooling channel also becomes optically thick. 

The introduction of even a trace amount of gas-phase metals alters this picture by increasing the cooling rate. Emission from semi-forbidden fine-structure lines [C\,II] and [O\,I] and from ro-vibrational lines of molecules such as H$_2$O, OH, and CO can be an effective facilitator of cooling early in the collapse, allowing the temperature to actually drop as the gas compresses. However, these cooling pathways lose their efficacy when the atomic or molecular level populations thermalize and the line cooling rate switches from a quadratic $\Lambda_{\rm line}\propto n^2$ to a linear scaling in density. Given that the scaling of the adiabatic compressional heating rate is super-linear $\Gamma_{\rm ad}\propto n^{3/2}$, gas tends to heat up after level thermalization. This effect is evident in the thermodynamic trajectories at metallicities $Z>10^{-3}\,\zsun$ in which the cooling phase is seen to last only briefly.

The next significant thermodynamic feature in the metal-enriched tracks in Figure \ref{fig:Omukai} arises from the presence of dust grains. Unlike gas particles, grains are capable of radiating thermally. They can thus serve as much more powerful cooling agents provided that collisions between gas particles and grains are sufficiently frequent to facilitate efficient gas-to-dust heat transfer.  At low metallicities and densities, gas and dust may not be thermally coupled and their temperatures may differ, $T\neq T_{\rm d}$. As the collapse proceeds and density increases, collisions between the warmer gas particles and the cooler dust grains increase in frequency and eventually thermal coupling is established. The second, higher-density temperature minimum in the $Z>10^{-6}\,\zsun$ thermodynamic trajectories in Figure \ref{fig:Omukai} reflects this thermal coupling.  The minimum occurs at a critical density that increases with decreasing metallicity. In practice, $T_{\rm d}$ is computed by taking dust grains to be in instantaneous thermal equilibrium ($dT_{\rm d}/dt=0$) in the presence of thermal emission, CMB photon absorption, and collisional energy transfer with gas \cite{Schneider06,SafranekShrader14a}. At high enough densities the collisional energy transfer dominates and gas is thermally coupled to dust.

Independent of metallicity, thermodynamic tracks converge to the same trajectory at high densities, $n\sim10^{11}-10^{16}\,\cc$ (increasing with decreasing metallicity).
 The common trajectory represents a collapsing cloud, one Jeans length in thickness, that is becoming opaque to continuum cooling provided by either dust thermal emission or $\htwo$ CIE.  It is this `opacity-limit' for fragmentation that sets the minimum mass of a gravitationally unstable cloud fragment \cite{Rees76,Low76}. Once the gas does become opaque to its own cooling radiation, the subsequent thermal evolution is adiabatic because radiation cannot escape and remove the internal energy generated by gravitational compression.

All of the mentioned thermal processes occur in lockstep with the non-equilibrium evolution of the chemical state. In metal-free chemical reaction networks, the $\htwo$ abundance remains strongly out of equilibrium, of the order of $x_{\htwo}\sim10^{-3}$, at least until three-body $\htwo$ formation kicks in around $n\sim10^8\,\cc$.   Recall that accurate computation of the $\htwo$ abundance was critically important for correctly modeling metal-free star formation. Now, even a trace of metals introduces additional complexity: dust grains catalyze $\htwo$ formation and metal-based molecules are synthesized through complex chemical reaction pathways with uncertain, temperature-sensitive reaction rates.

\begin{figure}[t]
\begin{center}
\includegraphics[width=0.75\textwidth]{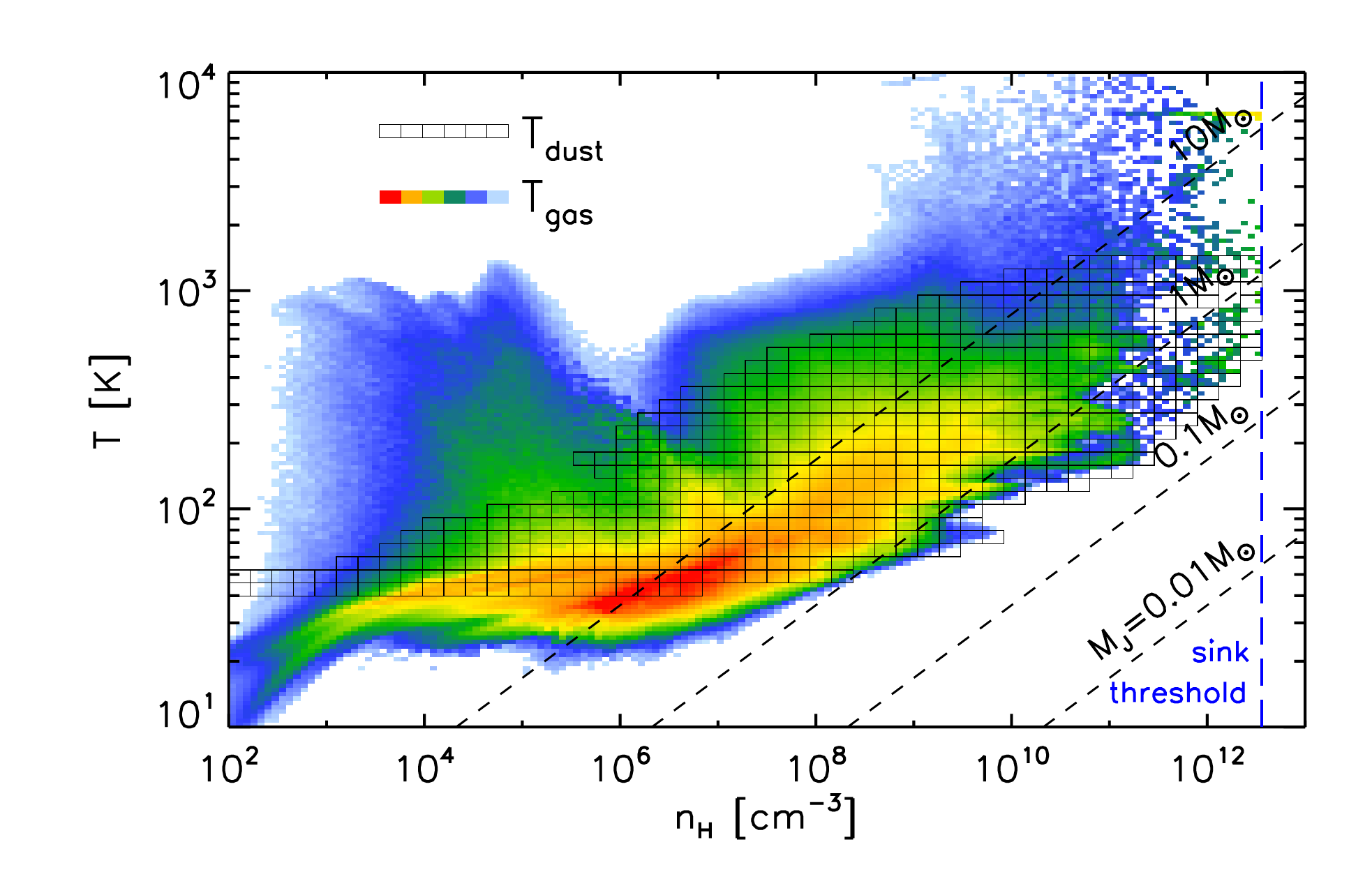}
\end{center}
\caption{Thermodynamic state of the gas and dust at the end of the simulation of Safranek-Shrader et al.\ \cite{SafranekShrader15}. Color represents the amount of gas in density-temperature cells with red representing the highest gas mass per cell. The dust temperature is overlaid with unfilled rectangles. Dashed lines indicate representative values of the Jeans mass. }
\label{fig:phase}    
\end{figure}

Numerical integration of such a chemo-thermodynamical network is computationally expensive. When embedded in a hydrodynamic simulation, it can dominate the computational cost.  A chemical reaction network constitutes an extremely stiff system of $N_{\rm spec}+1$ coupled ordinary differential equations (including one equation for gas temperature). A fully implicit approach to integrating such a system requires repeated solution of a $N_{\rm spec}+1$ dimensional linear system and such methods generally scale as $\mathcal{O}(N_{\rm spec}^3)$. A minimum of nine chemical species is generally required to describe the metal-free network active in the formation of metal-free stars (H, $\htwo$, e$^-$, H$^+$, H$^-$, H$_2^+$, He, He$^+$, He$^{++}$). Excluding metal-based molecules, about twice as many species are needed to provide a rudimentary model of a metal-enriched gas. The required number of species depends on a need to discern metal and  ionization states and is specific to the physical context at hand. Some species such as H$^-$ and H$_2^+$ are highly reactive and can be safely assumed to be in instantaneous chemical equilibrium; they can thus be treated as dependent variables \cite{Glover07}.  As an example of a cosmological hydrodynamical simulation of this kind, in Figure \ref{fig:phase} we show the thermodynamic state of gas and dust in Safranek-Shrader et al.\ \cite{SafranekShrader15}. The simulation tracked the atomic species C, O, and Si along with integrating a metal-free network and dust grain thermal and chemical processes.

Accurately computing the abundances of metallic molecules such as CO, H$_2$O, and OH requires many intermediary molecular and ionic species. These species are not only thermodynamically important but are  indispensable observational tracers of star forming gas. Including them can raise the number of chemical species in the reaction network to $N_{\rm spec}\gtrsim \mathcal{O}(10^2)$ \cite{Omukai05,Glover12} and integration of such a network can become computationally expensive.  One thus tries to be as parsimonious as possible in the selection of chemical species to track.\footnote{We believe, however, that future lies in the development of robust, comprehensive chemical solvers taking advantage of next generation hardware technologies and machine-learning algorithms.}  For example, Glover et al.\ \cite{Glover07} argue that for densities $n>(1\,{\rm Gyr}/t_{\rm sim})\,\cc$, where $t_{\rm sim}$ is the runtime of a simulation, CO and H$_2$O may dominate the cooling and should be included in the network. In contrast, Omukai et al.\ \cite{Omukai05} argue that neglecting metallic molecules has a minimal effect on the thermodynamic evolution of one-zone, collapsing prestellar cores.  Many other accuracy-preserving simplifications of chemical reaction networks have been proposed \cite{Nelson97,Glover12a}.

Metals can fundamentally alter the state of star-forming gas. In the Galaxy, stars seem to form exclusively in molecular gas. In neighboring galaxies, there is a well-established empirical correlation between the SFR and the surface density of molecular gas \cite{Leroy08,Blanc09,Schruba11,Kennicutt12,Leroy13}. This correlation is likely coincidental: there appears to be no \emph{a priori} reason that stars should form only in molecular, rather than more generally, in cold atomic gas.  Molecular hydrogen is not the principal gas cooling agent in star forming gas; C$^+$ and CO are far more potent coolants. 

The observed correlation between SFR and molecular hydrogen density seems to simply reflect an underlying correlation of the molecular hydrogen density and cold gas density.  It is the ability of gas to shield itself from FUV radiation that has the largest effect on the SFR.  The regions with high FUV extinction that favor $\htwo$ production also happen to have reduced rates of gas heating via the photoelectric effect on dust grains. Since the rate of $\htwo$ formation on dust grains and the degree of interstellar gas self-shielding from FUV radiation both scale with metallicity, one would expect the SFR-$\htwo$ correlation to also vary with metallicity.    Indeed, by conducting controlled numerical experiments with three-dimensional smoothed particle hydrodynamics (SPH) simulations, Glover \& Clark \cite{Glover12b} specifically showed that the presence of molecules does not determine whether gas can form stars.  At densities $<10^4\,\mathrm{cm}^{-3}$, gas efficiently cools by fine structure line emission from C$^+$ and at higher densities by transferring energy to dust that radiates thermally. 

At low metallicities ($\lesssim10^{-2}\,\zsun$), star formation is expected to occur before a significant fraction of hydrogen has been converted into molecular form.  Krumholz \cite{Krumholz12a} showed that at metallicities $\sim10^{-3}\,\zsun$ and standard ISM densities and temperatures, the time on which an initially atomic gas converts into molecular form is $\sim100$ times longer than the cooling time. This suggests that  low-metallicity clouds can cool and form stars, and likely be obliterated by stellar feedback, before any significant amount of $\htwo$ can form. 

Studies have incorporated the atomic-to-molecular transition into three-dimensional cosmological hydrodynamic simulations, but they typically operate on the imprecise assumption that star formation can occur only in molecular gas \cite{Gnedin09,Gnedin10,Gnedin11,Christensen12,Kuhlen12a,Kuhlen13}.  On the other hand, high resolution cosmological simulations that track metal enrichment but neglect dust-grain-catalyzed $\htwo$ formation \cite{Wise12,Wise12b} are incapable of forming molecular clouds.  When $\htwo$ is allowed to form on dust grains, there is indeed an atomic-to-molecular transition in early, metal-poor star forming objects, but only at densities well above $10^4\,\cc$ (e.g., $\sim 10^7\,\mathrm{cm}^{-3}$  \cite{SafranekShrader15}). It will be crucial for future studies to conclusively establish the precise physical parameters (e.g., metallicity, external radiation field strength, shielding column density) that determine the chemical state of star forming gas in the earliest galaxies.

\subsection{Metals and the Characteristic Stellar Mass}

In the local Universe, the observed stellar IMF $\xi(M)\propto M^{-\alpha}$ is an approximate power law with (negative) index $\alpha=2.35$ \cite{Salpeter55} at high masses $\gtrsim 1\,M_\odot$.  At lower masses it has an approximately log-normal form. Overall it exhibits remarkable uniformity over a wide range of star forming environments \cite{Bastian10}. The peak mass of the IMF where $d\xi/d\log M=0$ lies around a few tenths of a solar mass \cite{Kroupa02,Chabrier03} and is also fairly invariant across environments.  In the context of the transition toward metal-enriched star formation in the early Universe, we focus on the establishment and evolution of the peak mass.  The peak mass is interesting because the long-lived relic stars of this epoch in the Galactic neighborhood are close to the peak mass. Also, the question of the origin of the power-law tail remains comparatively less well settled.
 
Simulations \cite{Li03,Jappsen05} and theoretical considerations \cite{Larson05} support the hypothesis that by controlling thermodynamic evolution of gas, radiative heating and cooling govern fragmentation in gravitationally collapsing gas clumps.  In numerical experiments that idealize the equation of state with a polytropic relation $P\propto\rho^{\gamma}$, the degree of fragmentation strongly increases with decreasing polytropic index $\gamma$ (i.e., with the softening of the effective equation of state) \cite{Li03}.  With a non-polytropic equation of state, or if instead of parametrizing the equation of state one has directly integrated the thermal evolution equation coupled to a chemical reaction network, the characteristic fragmentation mass scale is approximately the Jeans mass evaluated at the last, highest-density minimum in the density-temperature evolutionary trajectory.  At this minimum the effective adiabatic index transitions from below to above unity \cite{Jappsen05}.

The idea that thermodynamic behavior governs fragmentation can be applied in one-zone gravitational collapse models to estimate the resulting fragment masses.  For $Z\geq 10^{-5}\,\zsun$, each thermodynamic track in Figure \ref{fig:Omukai} has two distinct temperature minima separated by $\sim 5$ dex in density. The minima can imprint characteristic fragmentation (and sub-fragmentation) mass scales. The lower density minimum is the saturation of cooling by fine-structure lines like C\,II and O\,I and the release of latent heat in H$_2$ formation. The higher-density temperature minimum reflects the cooling of gas by dust emission and the onset of gas-dust coupling. The characteristic fragmentation masses associated with the minima are the respective thermal Jeans masses evaluated at the density and temperature of each minimum.

If the characteristic stellar mass scale is indeed thermodynamically imprinted, then it seems that the formation of low mass stars generally requires dust.  Without dust, when metallic fine structure lines are the principal cooling agents, the minimum fragmentation mass varies non-monotonically with metallicity \cite{Bromm01,Omukai05,Santoro06,SafranekShrader10}. Such dust-free characteristic mass is too large to explain the formation of the low-mass, very metal poor stars found in the local Universe \cite{Klessen12}.  In three-dimensional simulations of cloud collapse, gas-dust coupling shifts the minimum fragmentation mass to the expected range, $\sim0.01-1\,\msun$ \cite{Clark08,Dopcke11,Dopcke13,SafranekShrader15}.  The metallicity required for dust-gas coupling to have significant thermodynamic impact and enable solar-mass-scale (or lower mass) fragmentation is rather low, $Z\sim10^{-6}-10^{-5}\,\zsun$ \cite{Omukai08,Chiaki14,Ji14,Chiaki15}.

\begin{figure}[t]
\begin{center}
\includegraphics[width=0.45\textwidth]{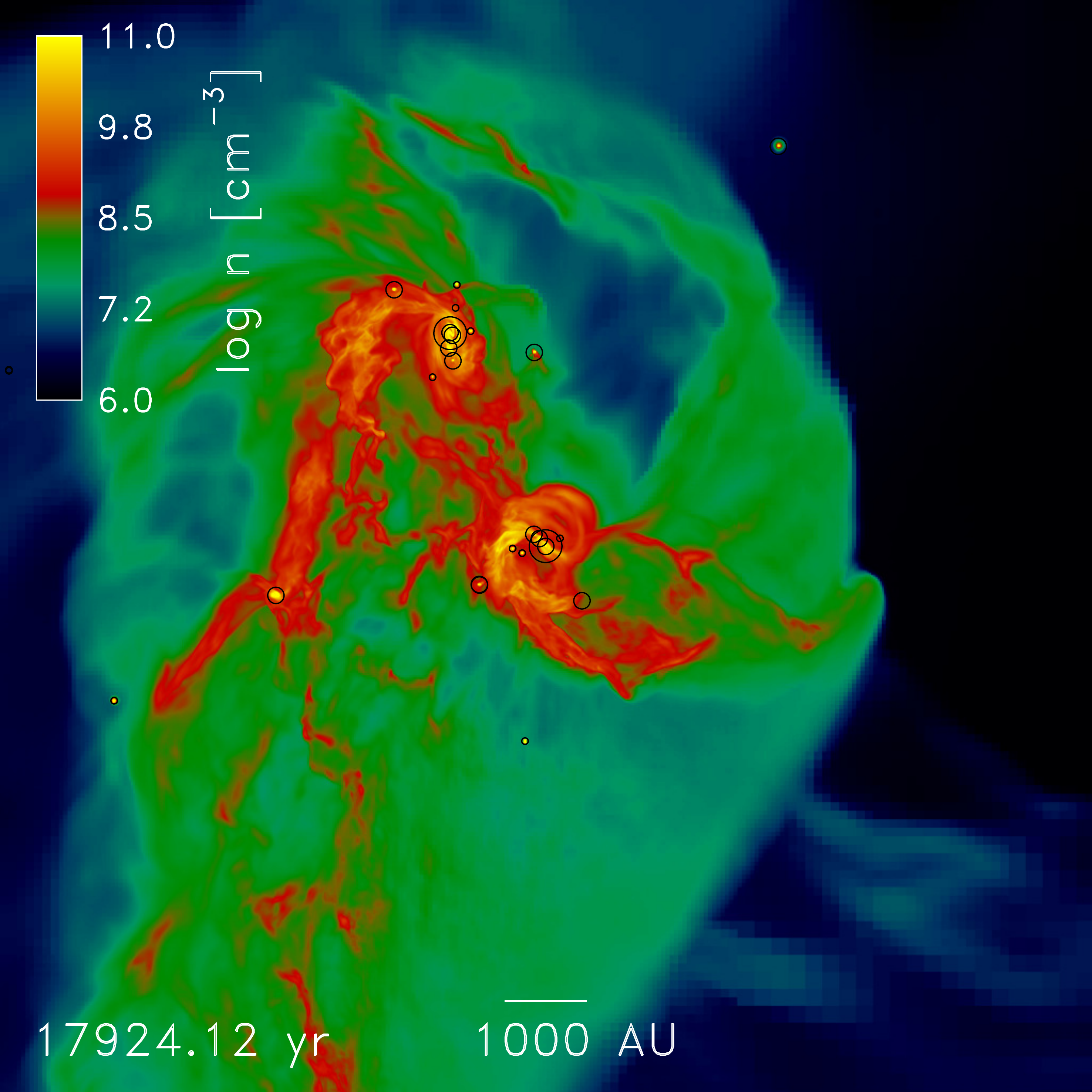}
\includegraphics[width=0.45\textwidth]{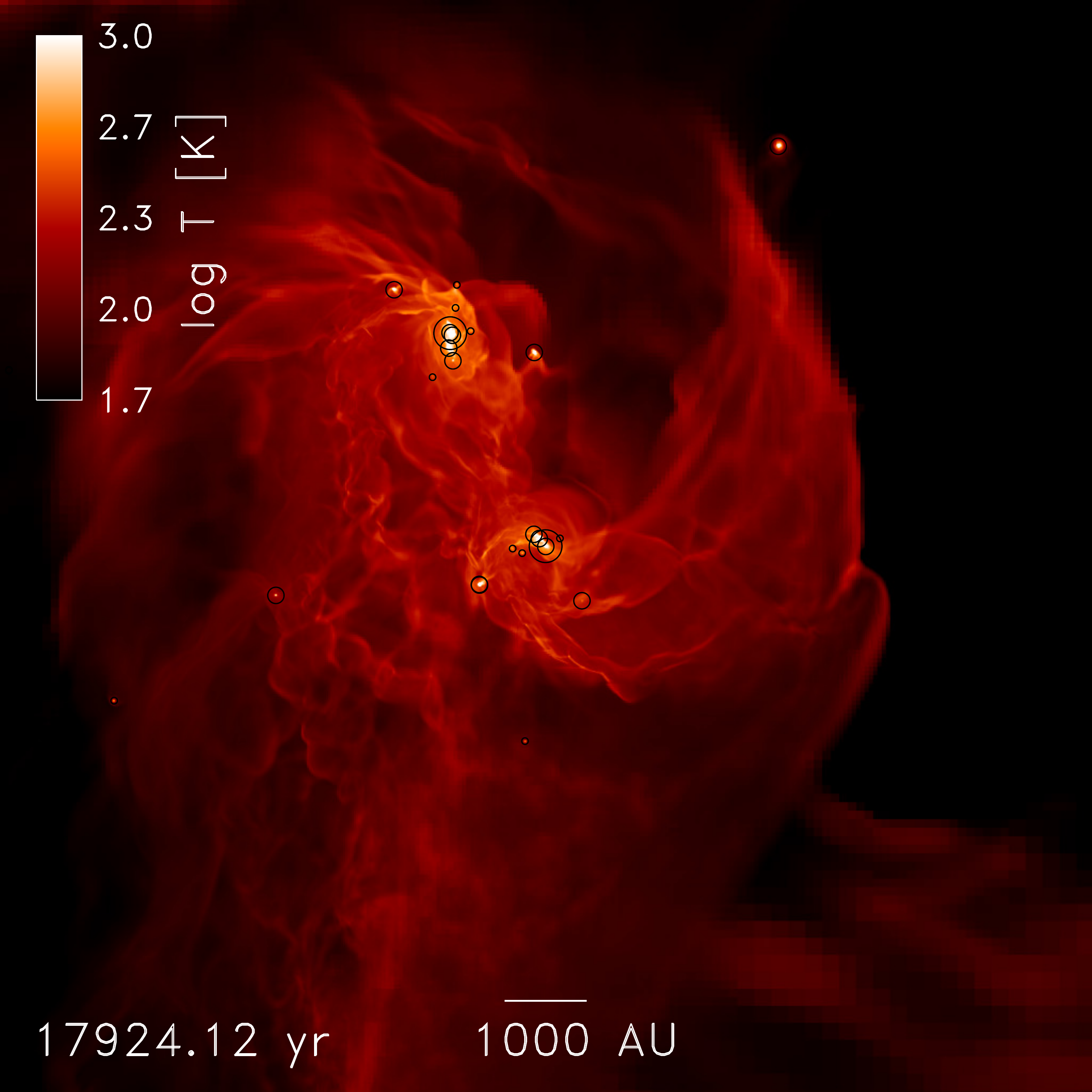}
\end{center}
\caption{Mass-weighted line-of-sight projections of gas density (left) and temperature (right) in the simulation of Safranek-Sharder et al.\ \cite{SafranekShrader15} that zooms-in on a pre-stellar clump in the parent cosmological simulation \cite{SafranekShrader14}. The host halo redshift and mass are $z=13.8$ and $M_{\rm 200}\approx 2\times10^7\,M_\odot$, respectively.  Circles mark projected  locations of sink particles representing protostars with the circle size increasing with sink mass. The lowest-mass sink $M<1\,\msun$ circles are drawn with twice the sink particle accretion radius $\racc=20\,\au$, the sinks with $1\,\msun<M<10\,\msun$ with $5\,\racc$, and the sinks with $M>10\,\msun$ with $10\,\racc$. Protostellar radiation raises the gas temperature by heating the dust.
}
\label{fig:dens_temp}    
\end{figure}

\begin{figure}[t]
\begin{center}
\includegraphics[width=0.5\textwidth]{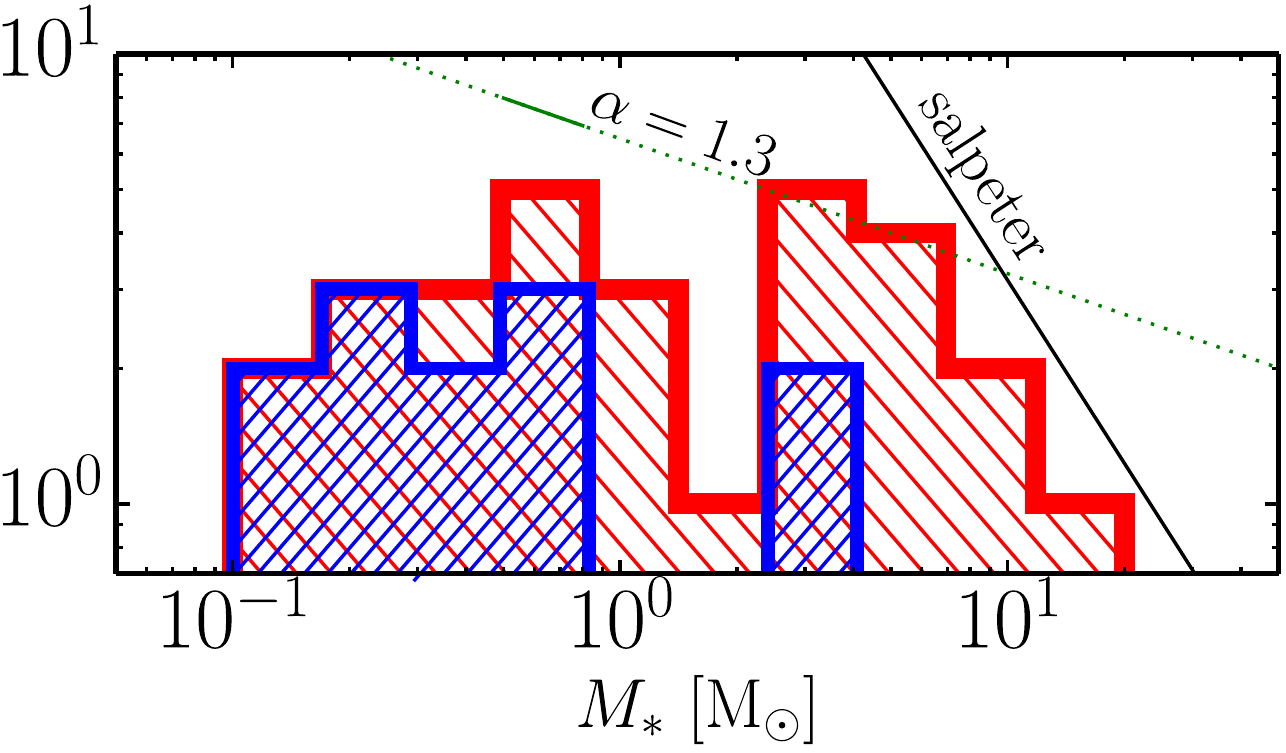}
\end{center}
\caption{Sink particle mass function at $18$ kyr after the onset of sink formation in the simulation of Safranek-Shrader et al.\ \cite{SafranekShrader15}. The red histograms are for all sink particles and the blue histograms are only for non-accreting sinks, namely those with instantaneous accretion rates below $10^{-7}\,\msunperyr$ in the last $500\,\mathrm{yr}$ of the simulation. The black and green straight lines indicate power-law slopes of Salpeter $\alpha=2.35$ \cite{Salpeter55} and Geha et al.\ $\alpha=1.3$ \cite{Geha13} where $\xi(M_*) \propto M_*^{-\alpha}$. The solid portion of the Geha et al.\ line, between $0.5\,M_\odot$ and $0.8\,\msun$, indicates the narrow stellar mass range in which observations constrain the IMF in two ultra-faint dwarf spheroidal satellites.
}
\label{fig:imf}    
\end{figure}

If the presence of dust in metal-enriched gas is assumed as a given (see \S~\ref{sec:dust} below), the formation of individual low-mass protostars can be simulated directly from cosmological density fluctuations, with the non-{\it ab-initio} part of the computation being nucleosynthesis and dust production. The results of the simulation of Safranek-Shrader et al.\ \cite{SafranekShrader14a,SafranekShrader15}, shown in Figure \ref{fig:dens_temp}, agree with one-zone models in that the gas passes through two distinct episodes of gravitational fragmentation, one facilitated by fine structure line cooling and the other by dust processes.  In the first episode, supersonic motions in a thermally unstable core with a characteristic density $\sim 10^4\,\cc$ excite density fluctuations that become self-gravitating. In the second episode, when density reaches $\sim10^8\,\cc$,  the interplay of dust cooling, the CMB temperature floor, and self-gravitating turbulence cascading from larger scales fragments the pre-stellar clumps into protostars with a characteristic mass of $\sim0.5\,\msun$.  The characteristic mass corresponds to the first rise of the protostellar IMF shown in Figure \ref{fig:imf}.  The simulation was run for only $\sim20\,\mathrm{kyr}$ after the formation of the first protostar in the parent pre-stellar clump.  This was long enough for massive, luminous protostars to form, heat the dust grains, and raise the local Jeans mass.  Since the star-forming system was not in a steady state of any kind, the characteristic mass may evolve on, say, $\sim 100\,\textrm{kyr}$ time scales.  Ionizing radiation from stars with masses $\gtrsim 10\,M_\odot$ can photoevaporate, prevent the fragmentation of, or even compress pre-stellar and star-forming clumps \cite{Krumholz06}. Or, perhaps, it may not have too inhibitive of an effect \cite{Peters10,Dale11}. To date, none of the simulations possessing sufficient mass and spatial resolution to identify the characteristic stellar mass has also included photoionization.  

With time, the star clusters forming in pre-stellar clumps created by the first fragmentation episode start interacting with each other,  by heating dust and photoionizing and photoevaporating gas throughout the star forming complex, and by interacting dynamically, merging, and virializing.  Direct numerical simulation of this process on $\sim1\,\mathrm{Myr}$ time scales is the looming challenge.  Substantial acceleration of the local chemical and thermodynamic update, if operator-split from the transport terms, can be attained by transferring the update to coprocessors in specialized microprocessor architectures such as the Intel Phi built on the Intel Many Integrated Core (MIC) Architecture model.  Radiative transfer algorithms adapted to star formation physics are also catching up \cite{Davis12,Klassen14,Tsang15}.  Somewhat farther from meeting the science requirements are algorithms for gravitational computations (Poisson solvers) on non-uniformly refined meshes with strongly-interacting fluid-like (gas) and point-like (dark matter and stars) gravitational sources.  Particularly challenging to simulate are few-body gravitational encounters and hierarchical systems (binaries, triples, etc.) that are generic in young star clusters.  Also, insufficient attention has been given to the robustness of the standard hyperbolic solvers (e.g., the Piecewise Parabolic Method for compressible inviscid hydrodynamics) in ionization and chemical fronts where thermodynamic source terms are confined to thin reaction zones.

\subsection{The Physics and Numerics of Metal Enrichment}
\label{sec:enrichment}

Just how does the star-forming medium become enriched with metals?  The abundance patterns in the most metal poor stars and intergalactic absorption systems require that nucleosynthesis-processed fluid elements be diluted with hydrogen-helium gas by a factor as high as $\sim 10^6$.  It suffices for the dilution to be completed, in a coarse-grained sense, down to stellar mass scales $\sim 0.1-1\,M_\odot$, but it need not be complete on the microscopic scales on which particle diffusion ultimately homogenizes the medium.  The numerical challenge is to model this dilution over its huge dynamic range. 

Any fluid motions that increase the surface area and geometric extent of the Lagrangian volume containing the yield of a nucleosynthetic event drive chemical dilution, again in a coarse-grained sense, on the spatial scales characterizing the motions. A continuously driven turbulent cascade facilitates dilution across a range of scales and can act as an efficient mixer \cite{Pan13}.  In many astrophysical environments, however, turbulence is driven intermittently rather than continuously.  Decaying turbulence from transient forcing by, e.g., shock waves, homogenizes nucleosynthetic products on coarse-grained mass scales while potentially not mixing the gas on finer scales \cite{Scalo04}.  Turbulence may enhance as well as reduce the homogeneity of heavy elements locked in dust particles \cite{Hopkins14b}.  Restricting the following discussion to ideal hydrodynamic instabilities we note that:

\begin{itemize}
\item Chemical dilution begins in the stellar interior, both in the hydrostatic and explosive phases.  Fluid-dynamical instabilities (Rayleigh-Taylor, Richtmyer-Meshkov, Kelvin-Helmholtz) excited by the supernova shock wave generate vorticity.  These and other instabilities can impart strong anisotropy to the explosions, ejecting some elements (e.g., the iron group) in a few discrete directions \cite{Wongwathanarat14}.  

\item The expanding supernova remnant is susceptible to vorticity-generating instabilities throughout its evolution \cite{Ritter12,Whalen13,Ritter14,Sluder15}.  Supernovae need not explode in isolation but can be clustered. Interactions between remnants provide strong momentum kicks to the heterogenous multiphase medium consisting of shocked supernova ejecta, shocked circumstellar medium, entrained dense clouds into which shocks do not penetrate, and thin radiative shells of evolved remnants.

\item Metal-enriched star formation is naturally preconditioned on gravitational collapse.  Simulations demonstrate collapse on all scales ranging from the cosmic web and the circumgalactic medium down to the scales of single pre-stellar clumps.  Certain stages of the collapse are supersonic and gravitational infall excites strong turbulence that homogenizes the star-forming gas.  Turbulent virialization \cite{Wise07,Prieto12} may in fact be responsible for the relative chemical homogeneity of compact stellar systems including globular clusters \cite{Feng14,Ritter14}. 

\end{itemize}

The ejecta of initially metal-free stars exploding in low-density H~II regions travel to radii $\sim1-10\,\mathrm{kpc}$ before reaching pressure equilibrium with the IGM \cite{Wise08,Greif10,Wise12,Ritter14}.  The vorticity of this protogalactic outflow is low, which implies that turbulence in the outflow is decaying, weak, and insufficient to drive metal dilution down to star forming scales.  The high-metallicity fluid contaminated with supernova ejecta remains confined in discrete pockets even as the outflow has turned around into an inflow and is re-collapsing in a grown dark matter halo. Eventually, however, as the gas virializes, the vortical time scale $|\nabla\times\mathbf{v}|^{-1}$ becomes short compared to the lifetime of the virialized system and dilution becomes possible.  For example, Ritter et al.\  \cite{Ritter14} find near-complete homogenization (drop of gas metallicity spread) at densities $\gtrsim 10\,\mathrm{cm}^{-3}$ in a recollapsing minihalo.

There is also the more extreme scenario where the supernova explodes in relatively dense gas. This can happen if the progenitor is relatively faint and the natal gas cloud is so dense, and perhaps in the state of continuous infall \cite{Whalen13}, that the H~II region does not break out.  It might also happen that a star about to go supernova engages in a strong, few body dynamical interaction with other stars and by chance gets slingshot into a nearby dense cloud.  When ejecta expand into dense gas, e.g., $n\gtrsim 100\,\mathrm{cm}^{-3}$, the free-free radiative cooling time can become shorter than the dynamical time in the reverse-shocked ejecta \cite{Kitayama05}.  The remnant transitions directly from free expansion into radiative cooling and rapidly condenses into a dense shell, thus skipping the adiabatic Sedov-Taylor regime altogether. While low-resolution or low-dimensional studies of this scenario exist \cite{Whalen13,Jeon14}, here we can only speculate about the detailed outcome of such explosions.  If the radiative shell remains quasi-spherical, the ejecta momentum gets deposited in $\sim10^3\,M_\odot$ of circumstellar gas before ambient pressure resists further expansion. This also means that the ejecta momentum does not get amplified by the large factor $\sqrt{M_{ST}/M_{\rm ej}}$, where $M_{\rm ej}$ is the ejecta mass and $M_{\rm ST}$ is to total shocked mass in the Sedov-Taylor phase, by which it is amplified if the energy-conserving phase does occur.  The promptly cooling remnant immediately enriches the gas in the vicinity of the explosion to $\gtrsim 0.1\,Z_\odot$.  If the shell develops fingers while still highly supersonic, then the ejecta may remain confined inside very metal rich, and likely chemically heterogeneous ``bullets" that can travel to large distances ($\sim0.1-1\,\mathrm{kpc}$) without becoming diluted in the ambient medium.

\begin{figure}[t]
\begin{center}
\includegraphics[width=\textwidth]{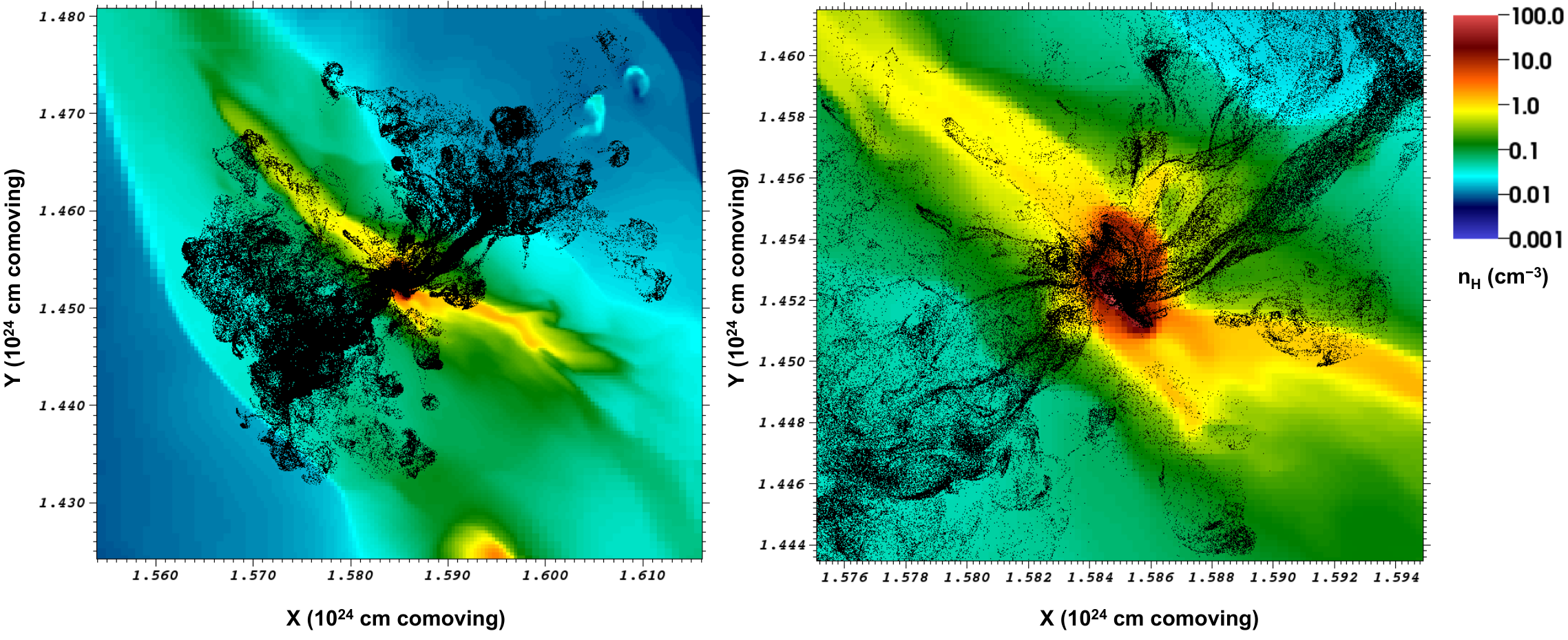}
\end{center}
\caption{Metal dispersal and fallback in a minihalo of Ritter et al.\ \cite{Ritter12}. The left panel is $1.1\,\mathrm{kpc}$ (physical) wide and centered on the gravitational potential minimum; the right panel is a $360\,\mathrm{pc}$ detail. The black points are the metal particles in projection and the color is a slice of the hydrogen density. Metal-carrying Rayleigh-Taylor fingers have a positive radial velocity and have breached the virial radius of the halo, while most of the metal mass is falling back into the halo center and remains incompletely mixed with the primordial gas.}
\label{fig:Ritter2012}
\end{figure}

\begin{figure}[t]
\begin{center}
\includegraphics[width=0.5\textwidth]{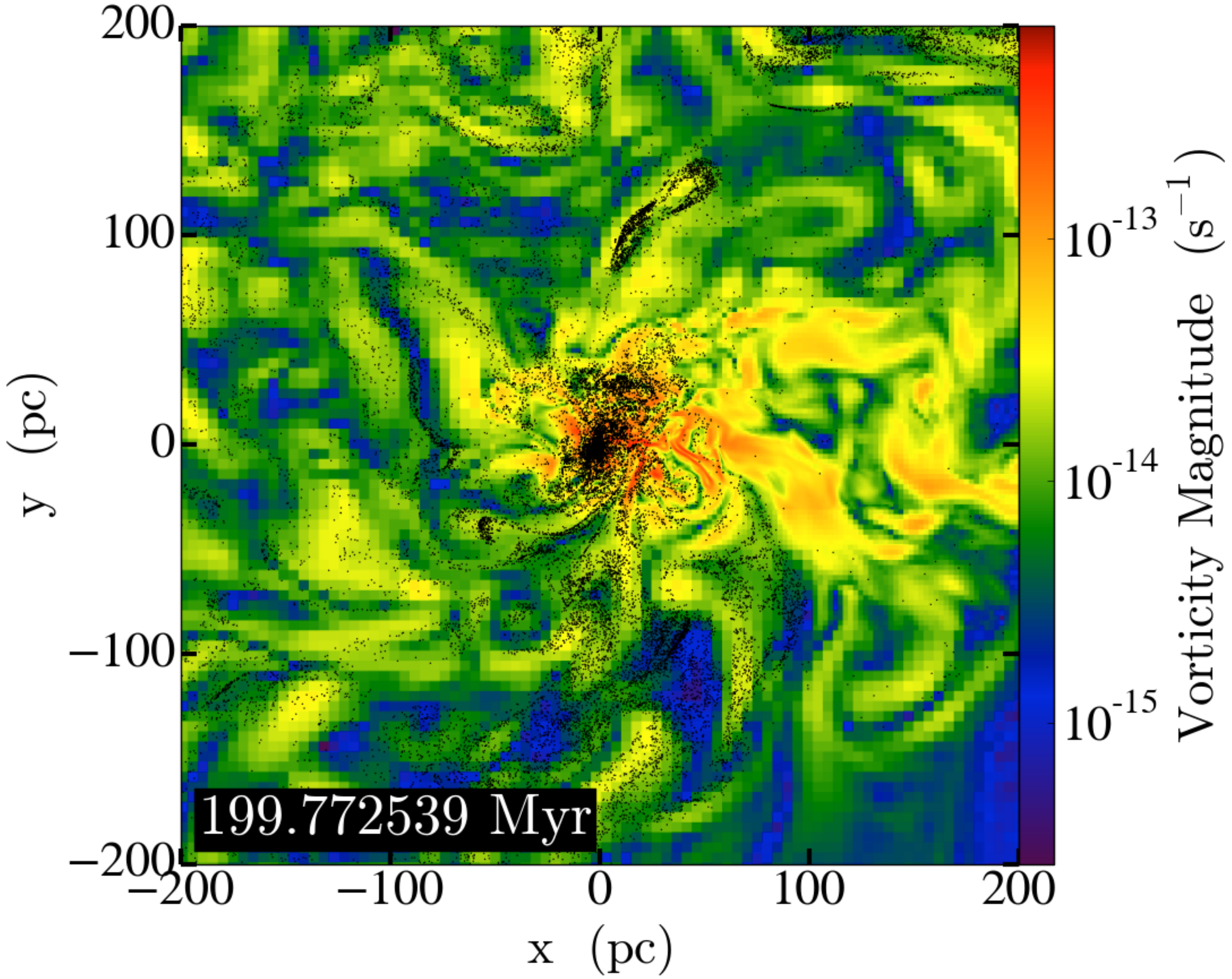}\hspace{0.75cm}
\includegraphics[width=0.4\textwidth]{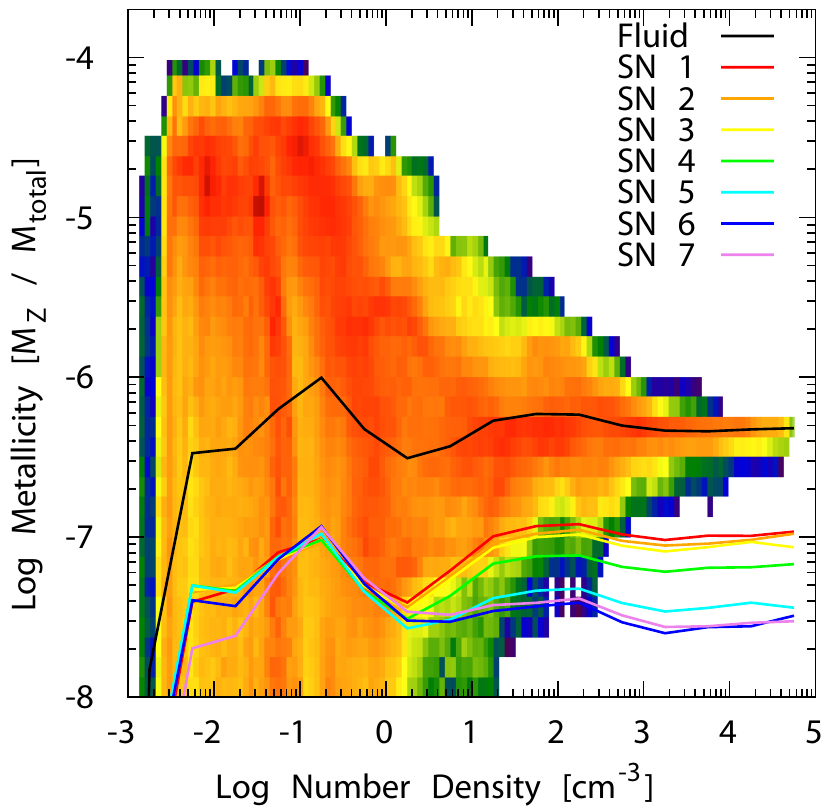}
\end{center}
\caption{Left: a slice of vorticity magnitude $|\nabla\times\mathbf{v}|$ in the center of a minihalo hosting seven consecutive core-collapse supernovae $200\,\mathrm{Myr}$ after the progenitor stars were inserted in the simulation of Ritter et al. \cite{Ritter14}.   The host halo redshift and mass are $z=11.7$ and $M_{\rm 200}\approx 2.5\times10^7\,M_\odot$, respectively.  The over-plotted black dots show Lagrangian particles tracing supernova ejecta in a $5\,\mathrm{pc}$ thick slab containing the slice. The metal ejecta are unmixed outside the central, tens of parsecs wide high-vorticity core. Right: metallicity as a function of gas density in the same simulation snapshot.  From red to blue, the color scales with the logarithm of the fluid mass in the density and metallicity bin. Solid curve is the mean metallicity and colored curves are the fractional contributions from the seven consecutive supernovae.  The dense gas poised to form new stars is under-abundant in the ejecta of the latest supernovae to explode in the same superbubble because the late ejecta are heated to higher entropy.}
\label{fig:Ritter2014}
\end{figure}

The geometry of the sheets and filaments of the cosmic web collimates the outflows powered by clustered core-collapse supernovae to expand into the flanking voids. In projection, the outflows can have ``butterfly'' or ``hourglass'' morphologies.  The outflows concentrate momentum into fingers and clumps and travel farther than a quasi-spherical model would predict (see Figure \ref{fig:Ritter2012} and Figure \ref{fig:Ritter2014}, left panel). The metals ejected from halos linger in the voids, while simultaneously, pristine (or at least much less metal rich) gas continues streaming into the same halos.  This means that the formation of extremely metal poor stars can continue for some time even in cosmic neighborhoods that have synthesized ample metal masses.  

The metallicities of second-generation stars may be more sensitive to the energies and the detailed explosion cadence of the first-generation supernovae than to the total metal yield.  Wimpy, isolated explosions can allow prompt fallback and enrichment to $Z\sim 0.01\,Z_\odot$, particularly in minihalos \cite{Ritter12,BlandHawthorn15}.  The metallicities of stars forming in the immediate wake of ultra-energetic explosions (PISNe) \cite{Cooke14} and concerted normal explosions \cite{Ritter14}---both delivering much higher metal masses than isolated explosions---can be substantially lower than after isolated explosions because the bulk of the metals are dispersed across longer cosmic wavelengths that take longer to collapse.  

Digging deeper into supernova remnant hydrodynamics, we discover effects that can skew the abundance patterns in metal-enriched stars away from the monolithic elemental yields of the contributing supernovae \cite{Ritter14,Sluder15}. One such effect concerns the dependence of the entropy of reverse-shocked ejecta on the mass coordinate in the explosion or the ejection velocity.  Reverse shock heating can leave the inner mass shells in isolated explosions at entropies too high to cool on halo dynamical times. In clustered, interacting explosions, the entropy depends on the temporal order of the explosion. There, the post-shock entropy increases with each explosion taking place inside the bubble blown by the preceding explosions.  This means that the abundances in enriched star-forming clouds can be deficient in the innermost mass shells of the contributing supernovae (if they exploded in isolation) and in the most recent supernovae to have exploded (if the explosions were clustered; Figure \ref{fig:Ritter2014}, right panel).  Another effect relates to angular anisotropy of radial deceleration as ejecta collide with an anisotropic distribution of dense clouds surrounding the explosion. This couples to any angular variation in the chemical composition of the ejecta, the variation that could arise from, e.g., Rayleigh-Taylor fingering preceding supernova shock breakout. Clearly, hydrodynamics gives rise to biases complicating the identification of nucleosynthetic sources in stellar chemical abundance spaces. The usually assumed monolithic mapping of source abundances onto the chemical fossil record can be broken.  Some of these effects may be the origin of the apparent abundance anomalies in the most metal-poor stars \cite{Feltzing09,Cohen13,Yong13}. They provide a potential explanation for the origin of carbon-enhanced, metal-poor stars (CEMPs) \cite{Cooke14,Ritter14,Sluder15}. 

High resolution ($\lesssim 1\,$pc) radiation-hydrodynamical simulations seeded from cosmological initial conditions that progress beyond the first instance of metal-enriched star formation are approaching halo masses $\sim10^9\,\msun$.  Such halos could be hosts to line-emitting protogalaxies potentially observable with JWST and thirty-meter-class ground telescopes \cite{Pawlik11,Pawlik13} and should be the systems that drove the early cosmic reionization.  However the price of the order-of-magnitude increase in halo mass and the relatively long galactic evolutionary time scales $\sim10^9\,\mathrm{yr}$ is paid in the inability to resolve individual pre-stellar clumps, the coarsest unambiguous star formation sites.  

Wise et al.\ \cite{Wise12,Wise14} simulated a $1\,{\rm Mpc}^3$ comoving box to redshift $z=7$.   In smoothly accreting halos not experiencing explosive starbursts, a balance of steady metal expulsion in outflows and metal-poor inflow from the cosmic web keeps the metallicity of newly forming stellar associations at $\sim10^{-2}\,\zsun$.   Stellar associations assembling in gas-rich major mergers of rapidly growing halos exhibit a much larger spread in metallicity. The convergence of these results in the peak gas density at which the star formation subgrid prescription is activated ($\sim 10^3\,\mathrm{cm}^{-3}$ \cite{Wise12}, a couple of orders of magnitude below the density of pre-stellar clumps \cite{SafranekShrader14}) and the resolution at which stellar feedback is implemented will undoubtedly be tested in the coming years.  It is already clear from this and other similar simulations that radiative feedback (in particular, photoionization heating and radiation pressure) are the key regulators of the assembly of the first galactic systems.

Simulating cosmic metal dilution is fraught with numerical pitfalls. The advection of passive mass scalars such as the metallicity $Z$ with finite-volume Eulerian methods is susceptible to severe unphysical numerical diffusion that introduces extended, exponential metallicity tails into metal-free clouds adjacent to metal enriched ones \cite{Plewa99}.  This can spuriously produce a metallicity floor under conditions in which turbulence is not vigorous enough to impart such a floor.   Since in the presence of dust, low mass star formation may be possible at metallicities as low as $Z_{\rm crit}\sim10^{-6}\,\zsun$ \cite{Schneider06,Omukai08}, numerical diffusion can mislead us about the character (e.g., the IMF) of star formation.  If supernova ejecta are redundantly tracked with Lagrangian passive tracer particles, then one often finds that the ejecta fluid as traced by the particles remains undiluted and confined to thin sheets and filaments separated by pristine gas even when the Eulerian passive scalar tracer has suffered substantial dilution \cite{Ritter12,Ritter14}.  Tracer particle methods, however, are not without insidious issues of their own. The finite numerical order of the hydrodynamic transport scheme can spuriously cluster Lagrangian tracers and amplify metallicity contrast, opposite from the artifacts seen in the passive scalars.  In view of these limitations, the level of local metal enrichment can be trusted only in if the Eulerian and Lagrangian predictors of the local metallicity agree.

\subsection{The Origin of Dust}
\label{sec:dust}

The picture of dust-enabled low-mass star formation in early galaxies is still schematic and far from complete. Progress toward explaining the frighteningly complex dust physics seems an essential precondition for understanding the metallicity-driven evolution of the character of star formation in the early universe \cite{Schneider10,Ji14,Chiaki15}. However, even the most basic questions remain unanswered.  At redshifts $z\gtrsim10$, the lifetime of the typical dust-producing AGB stars exceeds the Hubble time \cite{DiCriscienzo13}. This seems to leave core-collapse supernovae as the only theoretically viable dust producers in the early Universe. Do the composition and amounts of dust produced in the supernovae of metal-free stars resemble those produced in the Galactic disk?  (Probably not, which makes it hard to extrapolate local relations to the early Universe.)  In the nearby Universe, supernovae seem be producing large quantities of dust, specifically in the form of large dust grains that resist destruction in shocks \cite{Gall14,Lau15}. Also, there is growing indirect evidence for an efficient production of dust by supernovae in high-redshift submillimeter galaxies \cite{Michalowski10,Rowlands14}. Some supernova-produced dust, especially in the form of small grains, is destroyed in the reverse shock \cite{Nozawa07} and in other interstellar shocks, but the grains that clearly survive seem to be able to re-grow in the gas phase.  

As an illustration of the complexity of dust formation in supernovae and the challenges that must be overcome if dust masses are to be computed from first principles, consider the nucleation of amorphous carbon grains \cite{Cherchneff10}.  When a fluid element in expanding ejecta from a carbon-oxygen shell cools to $\sim4000\,\mathrm{K}$ (this is well before the reverse shock), atomic carbon can become incorporated in pure carbon chains if it is not oxidized into CO.  Therefore, carbon dust formation is possible in fluid elements in which C abundance exceeds O abundance.  But if some He is present (due to hydrodynamic mixing), then electrons Compton up-scattered by gamma-rays from $^{56}$Co decay ionize He, and He$^+$, with its large ionization potential, destroys small carbon chains (and also destroys CO).  Thus the abundance of nucleated amorphous carbon clusters is extremely sensitive to the relative abundances of C, O, He, and $^{56}$Co in the ejecta, and these are in turn sensitive to the hydrodynamics of pre-supernova stellar evolution as well as explosion.

\subsection{Star Formation in Magnetized Clouds}

The magnetization of star-forming clouds is another effect that can influence the properties of stellar populations.   The Universe is expected to have emerged from recombination practically unmagnetized \cite{Widrow12}.  In cosmic plasmas, a ``seed'' magnetic field can be generated, for example, by baroclinity (the Biermann battery mechanism), and also in collisionless shocks (by kinetic instabilities).  Stellar evolution is a more powerful dynamo: magnetic field is generated in protostellar disks, stellar interiors, and compact remnant accretion disks.  The field is ejected into the ISM in stellar mass loss, supernovae, and protostellar and compact object jets \cite{Heinz08}. Therefore, any ISM or IGM polluted with the output of preceding stellar generations is magnetized.  

The physics of magnetic field amplification by gas turbulence is similar to the physics of chemical dilution (see \S~\ref{sec:enrichment}).  The equation governing magnetic field evolution has the same form as the equation governing gas vorticity evolution if one formally replaces magnetic resistivity with viscosity.   Just like turbulence ultimately dilutes a locally injected contaminant, so though random stretching and folding it transforms an arbitrary initial, localized magnetic field into a pervasive fluctuating field with an average energy density that is a fraction of the turbulent kinetic energy density \cite{Federrath11b}.  Generically, we expect nucleosynthetic events---supernovae and stellar mass loss---to liberate \emph{magnetized} metals.  Turbulence-generated magnetic field can then affect metal-enriched star formation.\footnote{Turbulence-amplified magnetic field does, however, affect the character of turbulence and can under certain circumstances attenuate the rate of turbulent metal dilution \cite{Sur14}.}

The role of a pre-existing magnetic field in star formation has recently been investigated with three-dimensional numerical simulations \cite{Price08,Federrath12,Myers13,Myers14}. The simulations paint a picture in which the magnetic field gives rise to what is essentially a factor-of-two (or similar order) effect of the magnetic field: the characteristic stellar mass is somewhat higher, the SFR is somewhat lower, and the clustering of stars at formation is higher. The one notable exception where magnetization may drastically affect the star formation outcome is in producing a shallower (harder) power-law tail in the distribution of gravitationally unstable pre-stellar cores in the high-mass regime, similar to the tail observed in the Galactic stellar IMF \cite{Padoan07}.

\section{Implications for Reionization}

The ionizing photon budget of currently known galaxy populations comes short of explaining the reionization of the Universe. 
The measured electron scattering optical depth to the surface of last scattering derived from CMB polarization maps \cite{Planck15} requires extrapolating the galaxy luminosity function during the epoch of reionization to faint magnitudes $M_V \sim -14$ or even $\sim -13$ \cite{Bouwens12,Finkelstein12}.  This brings into focus the role of low-luminosity galaxies in cosmic reionization and the evolution of the Universe.
Diverse, complementary observational and theoretical investigations of this phenomenologically rich epoch have isolated effective parameters relating to star formation that influence reionization's key global observables, namely the electron scattering optical depth to last scattering, the Ly$\alpha$ scattering optical depth to high-redshift sources \cite{Bolton13,Jensen13,Dijkstra14b}, the kinetic Sunyaev-Zel'dovich fluctuations \cite{Reichardt12,Mesinger12}, and in the near future, of the evolution of and fluctuations in the 21 cm hydrogen spin flip temperature of the IGM \cite{Furlanetto06,Morales10,Pritchard12}.  The effective parameters entering the reionization's ``source term" (A.\ Lidz, this volume) are:

\begin{itemize}

\item The stellar IMF is the linchpin parameter that determines the relative abundances of massive stars that are the primary sources of ionizing photons. The IMF appears to be remarkably uniform under the Galactic star formation conditions, though there is tentative observational evidence that the IMF exhibits systematic variation between the Milky Way and its dwarf satellite galaxies \cite{Geha13}.  It is encouraging that {\it ab initio} simulations of star formation are, as we have seen, on the verge of being able to predict the stellar IMF from cosmological first principles.

\item The stellar metallicities influence the photospheric temperatures and ionizing luminosities. The number of ionizing photons produced per baryon converted into stars $N_\gamma$ depends on the metallicity as well as on the IMF.  Metallicity introduces a new dimension to the state space of protogalactic star-forming clouds. Variation in this new dimension makes the full statistics of small-scale metal enrichment difficult to track with present computational capabilities.  Stochasticity in the metallicity and in the dust content could drive significant stochasticity in the IMF.  

\item The SFE can be expressed in terms of the fraction $f_\star$ of gas in a halo that has been converted into stars.  Rather than an independent, constant parameter, the SFE should be physically related to the IMF and the metallicity. For example, top-heavy IMFs and metal-poor stars produce bigger, hotter, more overpressured H~II regions. Radiation pressure acceleration of the ISM \cite{Wise12b} becomes stronger in more metal-rich and dusty star forming clouds.  More generally, the SFE should depend on the halo mass, redshift, and the proximity of more massive halos, though with large stochastic variation \cite{Wise14,OShea15}.

\item The halo baryon fraction $f_{\rm b}$ quantifies the gas fractions retained in halos.  Streaming of baryons relative to dark matter \cite{Naoz13}, photoionization \cite{Sobacchi13}, and energetic supernova events (e.g., PISNe \cite{Greif10}) can reduce the baryon fractions in low mass halos far below the cosmic mean $\Omega_{\rm b}/\Omega_{\rm m}$.  The baryon fraction is a critical parameter because baryon depletion can have a disproportionally strong suppressing effect on the SFEs \cite{Milosavljevic14}.  Models of reionization often assume a redshift-dependent minimum halo mass $M_{\rm min}$ below which, in ionized regions, $f_{\rm b}$ is substantially reduced so that $f_{\rm b}f_\star\sim 0$.  The effect of reionization on the baryon fraction depends on the halo growth and local reionization history.
When ionization fronts sweep past mini-halos containing neutral gas, they expel the mini-halo gas by explosive photoevaporation \cite{Barkana99,Shapiro04,Iliev05}.  When dark matter halos with maximum circular velocities $\lesssim 20\,\mathrm{km}\,\mathrm{s}^{-1}$ (the halos with masses $\lesssim M_{\rm min}(z)$) collapse in ionized patches, gas pressure counteracts gas collapse at the turnaround (when a collapsing perturbation decouples from the Hubble flow) or somewhat later, but before virialization \cite{Simpson14,Okamoto08,Sobacchi13,Noh14}.  Observations suggest that the effect of supernovae and other forms of stellar feedback on $f_{\rm b}$ is strong \cite{Erb15}. However the physics behind the precise observationally-determined scaling and statistics of feedback-controlled baryon fractions in galaxies is not well understood at a quantitative level and remains model-dependent \cite{Simpson13,Kuhlen13,Hopkins14,Wise14,Ma15,Onorbe15,Wheeler15}.  For example, one expects that the severity of the feedback from supernovae decreases with the halo gravitational potential depth (at constant SFR) and simultaneously increases with the SFR (at constant potential depth). The severity of feedback is thus fixed through the competition of at least two stochastic, correlated variables (the SFR should be correlated with the halo mass growth rate and thus, indirectly, with the gravitational potential depth).

\item The star formation duty cycles, or more precisely, the full stochastic time-domain structure of the SFR, governs---possibly in a highly nonlinear, history-dependent fashion---the fraction of time galaxies are LyC emitters \cite{ForeroRomero13} and the fraction of the synthesized metal mass that is expelled in outflows. The duty cycles are set by the physics of feedback which should itself be sensitive to the character of the IMF and the entire enrichment history of the ISM.

\item The LyC and Ly$\alpha$ escape fractions quantify the probability that ionizing photons emitted by stars and Ly$\alpha$ photons emitted in recombinations reach the IGM. The escape fractions should be particularly sensitive to the duty cycles: more bursty star formation with a higher supernova rate might be better at making radiative escape channels, though a single, instantaneous burst may not be---owing to the shortness of the lifetime of massive stars compared to the time it takes for dense clouds to be photoevaporated and for the supernova bubble to bust out of the galaxy and into the IGM  \cite{Kimm14,Ma15,Paardekooper:15}. The LyC escape fraction $f_{\rm esc}$ is empirically poorly constrained at any cosmic epoch.

\item The stellar, supernova, and accreting remnant luminosities and spectra beyond LyC determine the backgrounds in high-energy, long mean free path photons that heat the IGM much more uniformly than UV ionizing radiation \cite{Power09,Mirabel11,McQuinn12,Power13}. The amplitude of hard backgrounds will be reflected in the sky-averaged \cite{Fialkov14,Fialkov14c,Mirocha14,Xu14,Yajima15} and fluctuating \cite{Mesinger13,Fialkov14b,Mesinger14,Pacucci14} $21\,\mathrm{cm}$ signal.  Metal-free stars emit hard UV photons and their direct radiative impact on the IGM may also be detected in next-generation $21\,\mathrm{cm}$ surveys \cite{Ahn12,Ahn15,Yajima15}. The local impact of hard sources such as high mass X-ray binaries (HMXBs) may be sub-dominant to other forms of feedback from star formation \cite{Jeon14b,Jeon15}.
These effects should clearly be sensitive not only to the IMF, but also to the stellar binary fraction and the details of post-main-sequence evolution.  All these, of course, are themselves very sensitive to metallicity.  Stellar population synthesis models \cite{Fragos13} and observational indicators \cite{Dray06,BasuZych13,Prestwich13} suggest that metal-poorer HMXB populations should produce substantially higher X-ray luminosities per unit stellar mass formed.
\end{itemize}

Theoretical (typically semi-analytical or semi-numerical) models of the buildup of dissociating and ionizing backgrounds frequently implement crude prescriptions expressing the above parameters, such as, universal (and often rather large) SFEs in neutral halos and universal halo mass thresholds for the retention of baryons in ionized halos \cite{Fialkov13,Fialkov14}.  The latest intermediate-resolution ($\sim10\,{\rm pc}$) simulations, however, point to strong context dependence (i.e., the dependence on halo mass, redshift, star formation and metal enrichment history, and the history of irradiation by UV and X-ray backgrounds) in the SFEs \cite{Wise14}.  The shape of the IMF is usually treated as one of several independent subgrid parameters.  Other parameters relating to the feedback efficiencies are either tuned separately, or are numerically determined by the IMF and a subgrid SFE, together with, say, other implicit tunable parameters such as the resolution at which energy, momentum, and radiative flux fed back in by star formation are being deposited \cite{Agertz13,Gnedin14}.  

We have argued that the ultra-small-scale physics of the IMF could ``inversely cascade'' to much larger scales.  It is difficult to imagine understanding the stochasticity of early galaxy formation without deciphering, from the first principles, the question of the stellar IMF and, more broadly, without compiling a detailed understanding of star formation in the earliest galaxies.  At the same time, fast, approximate, parametric approaches \cite{Mesinger07,Mesinger11,Zahn11,Majumdar14} are a very useful complement to first-principles simulations for acquiring qualitative intuition about the sensitivity of the observational signatures to the subgrid models assumed.  They can be used to constrain the general properties of star formation from large-scale observations, such as to conclude, for example, reionization is invariable a patchy process even with a strong X-ray contribution \cite{Mesinger13}, and that neither X-rays \cite{McQuinn12,Mesinger13}, nor the star forming in metal-free minihalos could ha that the stars forming in metal-free minihalos could  have by themselves reionized the Universe \cite{Visbal15}.

\begin{acknowledgement}
M.~M.\ acknowledges support by NSF grant AST-1413501 and in part by NSF grant PHYS-1066293 and the hospitality of the Aspen Center for Physics.
\end{acknowledgement}

\bibliographystyle{unsrt}
\bibliography{chapter_MM_CSS}{}

\end{document}